\documentclass[%
 reprint,
amsmath,amssymb,
aps,
pra,
]{revtex4-1}

\usepackage{graphicx}
\usepackage{dcolumn}
\usepackage{bm}
\usepackage{hyperref}

\begin{document}

\preprint{APS/123-QED}

\title{Theory of cavity-enhanced non-destructive detection of photonic qubits in a solid-state atomic ensemble}

\author{Sumit Goswami$^1$}
\author{Khabat Heshami$^{2,3}$}

\author{Christoph Simon$^1$}
\affiliation{$^1$Department of Physics and Astronomy and Institute for Quantum Science and Technology, University of Calgary, Calgary, Alberta, Canada T2N 1N4}
\affiliation{$^2$National Research Council of Canada, 100 Sussex Drive, Ottawa, Ontario, Canada K1A 0R6}

\affiliation{$^3$Department of Physics, University of Ottawa, 598 King Edward, Ottawa, Ontario, Canada K1N 6N5}

%
%
%
%
%
\begin{abstract}

Non-destructive detection of photonic qubits will enable important applications in photonic quantum information processing and quantum communications. Here, we present an approach based on a solid-state cavity containing an ensemble of rare-earth ions. First a probe pulse containing many photons is stored in the ensemble. Then a single signal photon, which represents a time-bin qubit,  imprints a phase on the ensemble that is due to the AC Stark effect. This phase does not depend on the exact timing of the signal photon, which makes the detection insensitive to the time-bin qubit state. Then the probe pulse is retrieved and its phase is detected via homodyne detection. We show that the cavity leads to a dependence of the imprinted phase on the {\it probe} photon number, which leads to a spreading of the probe phase, in contrast to the simple shift that occurs in the absence of a cavity. However, we show that this scenario still allows non-destructive detection of the signal. We discuss potential implementations of the scheme, showing that high success probability and low loss should be simultaneously
achievable.

\end{abstract}
%
\maketitle


\section{Introduction}

The ability to non-destructively detect photonic qubits - without absorbing the photon and without revealing its qubit state - would enable important applications in photonic quantum information processing~\cite{nemoto2004nearly} and quantum networks~\cite{boone2015entanglement,simon2017towards}. One promising avenue towards this goal is quantum non-linear optics ~\cite{chang2014quantum}. Significant advances have been made through strong nonlinear interactions in atom-cavity systems~\cite{reiserer2013nondestructive}, nonlinearities mediated by Rydberg atoms ~\cite{saffman2010quantum} and AC Stark shift~\cite{schmidt1996giant,hosseini2016partially,feizpour2015observation}.

Recent progress in cavity-enhanced light-matter interfaces involving rare-earth ions (REI) succeeded in solid-state implementation of quantum memories and controlled light-matter interaction in single or ensembles of REIs doped into a crystal~\cite{zhong2017nanophotonic,zhong2017interfacing,faraon2018controlling,jeff2017isolating,hunger2018cavity}. This promises a path towards robust and scalable implementations of photonic quantum information processing. Driven by progress in coupling REI to nano-photonic cavities, a proposal for non-destructive photon detection based on a single REI coupled to a photonic cavity has been developed~\cite{o2016nondestructive}. However, at the current state of technology it is still challenging to achieve situations where a single ion is coupled to a cavity in a reproducible and scalable way. For practical reasons, it is therefore also of interest to consider employing REI {\it ensembles} in photonic cavities for non-destructive detection of photonic qubits.

One form of nonlinear interactions based on atomic ensembles is to use a single photon to
impart a detectable cross-phase shift on a multi-photon coherent probe field ~\cite{chen2013all,PhysRevLett.112.073901}. The simultaneous presence of signal and probe fields in different configurations in an atomic system enables cross-phase modulation based on the AC Stark shift~\cite{schmidt1996giant,venkataraman2013phase}. This effect is sensitive to the spatio-temporal overlap of probe and signal fields. Storing the probe field in the atomic ensemble eliminates any sensitivity to the timing of the signal~\cite{heshami2016quantum}, which can be exploited for non-destructive detection of photonic time-bin qubits without revealing any information about the time-bin state of the signal, as proposed in Ref. ~\cite{sinclair2016proposal}. Single-shot and non-destructive detection of single photons based on AC stark shift was shown to be impossible for a single pass through atomic ensembles, as off-resonant absorption loss becomes prohibitive for cross-phase shifts larger than the intrinsic phase uncertainty of the probe field~\cite{sinclair2016proposal}. This limitation can be circumvented with multiple passes through the medium~\cite{sinclair2016proposal} or by enhancing the cross-phase shift with a cavity. The multi-pass approach is difficult to realize in practice because it requires very low-loss switches. On the other hand, as we will see below, the cavity introduces complications that were not analyzed in Ref. \cite{sinclair2016proposal}, motivating the present study.

\begin{figure}[h]
\centering
\includegraphics*[viewport=40 490 1000 850, scale=0.49]{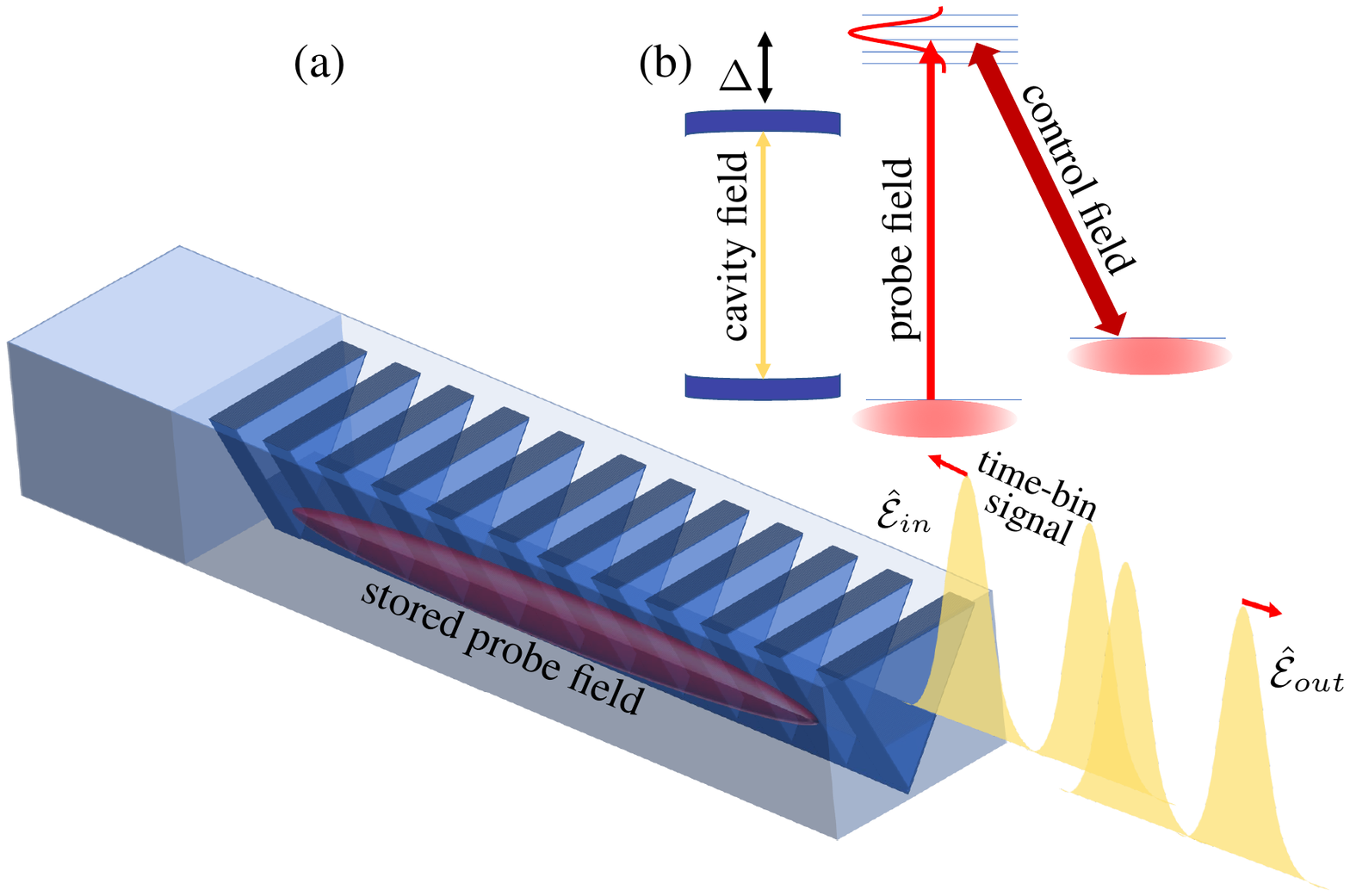}
\caption{(Color online) Scheme for non-destructive detection of a photonic time-bin qubit. (a) Input and output single-photon time-bin signal fields interacting with a nanophotonic crystal cavity coupled to an atomic ensemble that contains a stored probe field. (b) Atomic level configuration for storage of the probe field in the atomic ensemble. The cavity is in resonance with the probe for the storage process, but is detuned from the probe and brought into resonance with the signal for the non-destructive detection.}
\label{scheme}
\end{figure}

In this paper, we analyze a scheme to construct a single photon QND detector in a solid-state REI ensemble inside a cavity. A probe field is initially stored in the atomic ensemble. Then a single-photon signal that is resonant with the cavity and off-resonant with respect to the atomic transition interacts with the atomic ensemble; see Fig.~\ref{scheme}. Due to the AC Stark shift~\cite{autler1955stark}, a phase is imparted on the state of the atomic ensemble that contains a stored probe field. The phase shift on the atomic state leads to a phase shift on the retrieved probe field. In the absence of a cavity \cite{sinclair2016proposal}, an initial coherent state of the probe field $|\alpha\rangle$ is retrieved as $|\alpha e^{i\phi}\rangle$, where the signal pulse induces a phase shift $\phi$ on the probe state. For multiple passes this phase shift is simply multiplied by the number of passes. In contrast, in the current scheme, we find that the cavity resonance becomes sensitive to the number of atoms in the ground state, which depends on the photon number distribution of the stored probe field. This leads to a dependence of the cross-phase shift (due to the single-photon signal) on the photon number distribution of the probe field.

This paper is organized as follows. After discussing the storage of the probe field in section II, we analyze the cross-phase shift in detail in section III. In section IV, we show our results for a practical discrimination between a single photon and vacuum through quadrature detection. In section V, we describe an implementation based on rare-earth ion ensembles in nano-photonic cavities. In section VI, we conclude that the implementation of non-destructive photonic qubit detection should be within reach for the present approach.

\section{Probe storage}\label{probe_sec}

The proposal has two almost independent parts. First, it needs to be ensured that the probe pulse, in a many photon coherent state, can be efficiently stored and retrieved from the ensemble. Next comes the consideration that the signal, without being absorbed, can give enough phase shifts to the atomic ensemble storing the probe that the retrieved probe state can be distinguished from the initial probe state.

In section \ref{phase_section} we focus on the the phase shift due to the signal on the probe. However, to have that effect we need to store and retrieve the probe pulse efficiently. Light storage in quantum memories has been demonstrated in single photon level in variety of systems \citep{yang2016efficient,afzelius2009multimode,de2008solid,hedges2010efficient,zhong2015optically}. Specifically, single photon storage has been demonstrated conclusively in rare-earth ions doped crystals using Atomic Frequency Comb (AFC) quantum memory protocol with the added advantage of multimode storage\citep{afzelius2009multimode,de2008solid}. With our probe pulse, we are aiming to store a relatively intense coherent pulse which is in principle simpler than storing a single photon. The only significant difficulty regarding storing an intense pulse is that the number of photons stored should be much lower than the total number of atoms participating in the storage. Otherwise a significant portion of the atoms will reach the excited state during storage which will violate the assumptions of the standard linear quantum memory storage protocols \cite{gorshkov2007universal}. In Section \ref{Implement_sec} we provided numerical estimates for parameters of our proposal where we ensured such a parameter regime.

One of the fundamental constraints for probe storage is governed by the phase shift requirement of our protocol. As we are using an optical cavity, the cavity transverse area must be small to enhance the signal electric field considerably for a large phase shift. Due to this exact same reason of higher lateral confinement to increase the phase shift, waveguides were used in \citep{sinclair2016proposal}.

Storage efficiency of quantum memories increases with increasing optical depth \cite{gorshkov2007universal} and optical depth is proportional to both density and length of the medium (here, rare earth ions doped inside cavity) light is stored. In case of probe storage without a cavity, as in the waveguide case of \citep{sinclair2016proposal}, storage pulse can be incident along the same direction (e.g. along the waveguide in \citep{sinclair2016proposal}) as the signal. But in presence of a cavity the probe cannot be stored directly along the cavity if the signal and probe are detuned. This is because to maintain low signal loss we must detune the cavity from the atoms so that spontaneous emission is not enhanced. Hence, if the probe needs to be on resonance with the atoms for storage (as in the AFC protocol), it is not in resonance with the cavity anymore. As an alternative one can try to store the probe from the side of the cavity. However in that case, as we need a small transverse area cavity for large phase shift, the transverse length of the cavity is very small resulting in extremely small optical depth. To overcome this problem, the probe pulse must be stored along the cavity where there is enough optical depth and the cavity also enhances the storage efficiency significantly. In the AFC protocol the probe needs to be on resonance with the atoms. Hence to do probe storage along the cavity we need to have the cavity on resonance with the probe (and so with the atoms too) while the probe is stored. However, when the signal arrives, we need the cavity to be on resonance with the signal and detuned from the atoms to keep signal loss minimal. To solve this problem, we can dynamically control the cavity resonance frequency so that it is in resonance with the probe for probe storage and retrieval, but in resonance with the signal for non-destructive detection. This is feasible with current technology of piezoelectric motion controllers as we only need to detune the cavity a few picometers within a time span of around microsecond storage times of AFC memory. Requirements for the piezo-electric motion controller are discussed in more detail later, once we estimate the system parameters needed for implementation in section \ref{Implement_sec}. About the storage itself, AFC quantum memory protocol has been demonstrated to implement a high efficiency (56 $\%$) quantum storage in rare-earth ion doped crystal inside a cavity \citep{sabooni2013efficient}.  So, with the addition of the dynamical detuning, AFC protocol is one definite way to store the probe efficiently.

Another approach will be to keep the cavity permanently detuned from the atoms. But the probe needs to be on resonance to the cavity to be stored efficiently. Hence, we can implement an off-resonant Raman storage protocol. Although a Raman memory has not been demonstrated in the rare-earth ion doped crystal yet, it has been demonstrated widely in atomic gases \citep{saunders2016cavity,reim2010towards}. With the recent advancement in fabricating high finesse nano-photonic cavities \citep{zhong2015nanophotonic,zhong2016high,zhong2015optically,zhong2017nanophotonic, zhong2017interfacing} and stoichiometric crystals \citep{ahlefeldt2013precision,ahlefeldt2016ultranarrow} implementation of a Raman memory storage in rare-earth ions seems well within reach.

Concerning the storage state, the probe pulse can either be stored in the excited state or in a second ground state. However, it is preferable to store the probe in a second ground state for multiple reasons. First of all, the memory lifetime will then be limited by spin coherence time of the second ground state instead of the much shorter optical coherence time of the excited state. This will provide more time for the signal to pass and also for the dynamical detuning of the cavity. As the photon number in the probe pulse is a fraction of the total number of atoms, the second ground state contain much less number of atoms compared to the original ground state. So, if the signal photon imparts the phase shift on the second ground state where the probe pulse is stored the loss will be less. Another significant issue will be to store the probe pulse in the excited state will be the probability of stimulated emission while the signal is passing from excited state atoms in which the probe is stored. This may affect the signal fidelity. Hence, storing the probe in a second ground state will definitely be preferred if possible in a particular system. However, it may not be feasible for all systems. It depends on how many ground states are there in the particular system (rare earth ion). If only two ground states are used it may not be feasible for all protocols as the other ground state may be used for optical pumping to prepare a quantum memory. This is what constrained us in our example in Section \ref{Implement_sec} where we used an AFC quantum memory in Nd:YVO. Hence we considered storing in the excited state for this particular example.

\section{Cross-phase shift in the cavity}\label{phase_section}

\subsection{Theoretical model for cavity-enhanced QND}

In our proposal, once the probe is stored in the atomic ensemble inside the cavity a signal detuned from the atoms passes through the cavity inducing a phase shift on the atomic ensemble through the AC stark effect. A theoretical model is constructed for the phase shift that a signal photon induces on the atoms following~\cite{sinclair2016proposal}. However, our proposal deviates from~\cite{sinclair2016proposal} in that the phase shift now occurs inside a cavity. As we will see later, cavity modifies the phase shift depending on the number of stored probe photons which we modeled by calculating the cavity field and its interaction with the atoms. Following \citep{sinclair2016proposal} we start the theory by formulating the total Hamiltonian that governs our proposed system of signal and atomic ensemble.

\begin{eqnarray}\label{H_tot}
&{\hat H}_{tot}={\hat H}_0+{\hat H}_{int},\\
&{\hat H}_0=\hbar\omega_s a^{\dagger}a+\sum_{\delta}\hbar(\omega_{ge}+\delta)N(\delta){\hat \sigma}_{ee}\\
\label{H_int}
&{\hat H}_{int}=-\hbar g\left[ {\hat {\mathcal E}}e^{i\Delta t} \sum_{\delta}N(\delta) {\hat \sigma}_{eg} (t;\delta) +H.C.\right],
\end{eqnarray}
where the cavity field ${\hat {\mathcal E}}={\hat a}e^{i\omega_s t}$.

The Hamiltonians are written in terms of collective atomic operators defined as follows
\begin{equation}
{\hat\sigma}_{\nu\nu}(t;\delta)=\frac{1}{N(\delta)}\sum_{i=1}^{N(\delta)} {\hat\sigma}_{\nu\nu}^{i}(t;\delta); \nu=\{g,e\},
\end{equation}
and
\begin{equation}\label{dyn4}
{\hat\sigma}_{eg}(t;\delta)=\frac{1}{N(\delta)}\sum_{i=1}^{N(\delta)} {\hat\sigma}_{eg}^{i}(t;\delta)e^{-i\omega_p(t-z_i/c)}.
\end{equation}
where, individual atomic operators for the $j^{\text{th}}$ atom at position $z_j$ are given by ${\hat\sigma}^j_{\nu\nu'}=|\nu\rangle^j\langle\nu'|$, with $\nu,\nu'=\{g,e\}$. N($\delta$) is the number of atoms in frequency mode $\delta$ where the detuning of this particular mode from the central frequency is given by $\delta$ . The atomic ensemble has a central frequency given by $\omega_{eg}$ while the cavity, on resonance with the signal, is detuned by an amount $\Delta$ from the atoms and has a frequency $\omega_s$.

For relatively large detuning $\Delta$, we find an effective interaction Hamiltonian to describe the dynamics of the atomic polarization due to off-resonant interaction with the cavity field ${\hat {\mathcal E}}$. We start by finding the dynamics of the collective atomic operator in the Heisenberg picture

\begin{eqnarray}
&&{\dot{\hat \sigma}}_{eg}(t;\delta)=\frac{i}{\hbar}\left[ {\hat H}_{int}, {\hat \sigma}_{eg}(t;\delta)\right] \\
&& = -ig{\hat {\mathcal E}}^{\dagger} e^{-i\Delta t}\left( {\hat \sigma}_{gg}(t;\delta)-{\hat \sigma}_{ee}(t;\delta)\right).
\end{eqnarray}

This leads to

\begin{equation}
{\hat \sigma}_{eg}(t;\delta)= -i g \int_{0}^{t} dt' e^{-i\Delta t'}{\hat {\mathcal E}}^{\dagger}_s(t')\left({\hat \sigma}_{gg}(t';\delta)-{\hat \sigma}_{ee}(t';\delta) \right).
\end{equation}

If the signal passes for a time interval $\tau_s$ signal bandwidth is given by $1/\tau_s$. Under the approximation of large detuning ($\Delta >> 1/\tau_s$) compared to signal bandwidth, for any signal field shape this integral can be evaluated approximately by first integrating over the fast varying part $e^{-i\Delta t'}$ then multiplying it by the final value of the rest of the slow varying part.

\begin{eqnarray}\label{sig}
{\hat \sigma}_{eg}(t;\delta)=\frac{g}{\Delta} e^{-i\Delta t}{\hat {\mathcal E}}^{\dagger}_s(t) \left({\hat \sigma}_{gg}(t;\delta)-{\hat \sigma}_{ee}(t;\delta) \right).
\label{sigma_eg}
\end{eqnarray}

An effective Hamiltonian of the following form can be deduced from Eq. (\ref{H_int}) using Eq. (\ref{sigma_eg}).

\begin{equation}
{\hat H}_{int}^{eff}=-\frac{\hbar g^2}{\Delta} \sum_{\delta} N(\delta)\left( {\hat {\mathcal E}}{\hat {\mathcal E}}^{\dagger}+{\hat {\mathcal E}}^{\dagger}{\hat {\mathcal E}}\right)({\hat \sigma}_{gg}(t;\delta)-{\hat \sigma}_{ee}(t;\delta)).
\end{equation}


Using the Heisenberg relation we can find the dynamics of the atomic polarization using the free evolution and the above effective interaction Hamiltonians;

\begin{multline}
{\dot {\hat \sigma}}_{eg}(z,t;\delta)=i\delta {\hat \sigma}_{eg}(z,t;\delta)\\
+\frac{2ig^2}{\Delta} \left({\hat {\mathcal E}}_s(z,t){\hat {\mathcal E}}^{\dagger}_s(z,t)+ H.c.\right)  {\hat \sigma}_{eg}(z,t;\delta).
\end{multline}
This can be used to calculate the phase shift on atoms due to the signal field.
\begin{equation}
{\hat \sigma}_{eg}(t=T_2;\delta)=e^{i\delta t} e^{i {\hat \Phi}}{\hat \sigma}_{eg}(t=T_1;\delta),
\end{equation}
where,
\begin{equation}
{\hat \Phi}=\int_{T_1}^{T_2}dt' \frac{2g^2}{\Delta}\left( {\hat {\mathcal E}}{\hat {\mathcal E}}^{\dagger}+{\hat {\mathcal E}}^{\dagger}{\hat {\mathcal E}}\right)
\label{Phi_eq}
\end{equation}

Up to this point, we simply found the phase shift a signal will induce while passing off-resonant to an atomic ensemble in a cavity. The above consideration is fairly general in that it does not assume anything about the system. This treatment will be valid for atomic ensembles in a cold gas or a solid state system for a propagating signal or a cavity field. The difference between the propagating \citep{sinclair2016proposal} (waveguide or free space) and the cavity case lies in the electric field operator $\hat {\mathcal E}$ that we need to put in Eq. (\ref{Phi_eq}) in order to find the phase shift. In our proposal, inside a cavity the electric field gets changed from the free space case due to the atom-cavity interaction which will play a pivotal role in our analysis. Here, we derive the cavity field $\hat {\mathcal E}$ based on its dynamics and the cavity input-output relation \citep{scully1999quantum}, where we introduce the input signal field $\hat {\mathcal E}_{in}$. The rate of change in cavity field $\hat {\mathcal E}$ is given by
\begin{equation}
{\dot {\hat {\mathcal E}}}(t)= -\kappa {\hat {\mathcal E}}(t)+\sqrt{2\kappa}{\hat {\mathcal E}}_{in}(t)+\frac{2ig^2}{\Delta}({\hat \sigma}_{gg}-{\hat \sigma}_{ee}){\hat {\mathcal E}}(t).
\end{equation}

A probe pulse, which is stored into the atomic memory, is in a many-photon coherent state with an average photon number $N_p$. We assume that the probe pulse is stored into a different ground state i.e. ${\hat \sigma}_{ee}$ = 0. Hence, there are $N_g$ number of atoms in the ground state $ ({\hat \sigma}_{gg}-{\hat \sigma}_{ee}) = N_g$, resulting in
\begin{equation}
{\dot {\hat {\mathcal E}}}(t)= -\kappa {\hat {\mathcal E}}(t)+\sqrt{2\kappa}{\hat {\mathcal E}}_{in}(t)+\frac{2ig^2}{\Delta}N_g{\hat {\mathcal E}}(t).
\end{equation}
For a cavity with a relatively fast decay rate compared to the duration of the input signal field, a steady state solution can be used to describe the cavity field,
\begin{equation}
{\hat {\mathcal E}}(t)=\frac{\sqrt{2\kappa}}{\kappa-\frac{2iN_gg^2}{\Delta}}{\hat {\mathcal E}}_{in}(t).
\label{E_t}
\end{equation}

This enables us to find the phase shift per signal photon in the next step. Here, we consider a situation with a fixed number of atoms ($N_g$) in the ground state. The rather complicated scenario of our proposal where the many photon probe pulse (in a coherent state) is stored into the atoms before the signal arrives is not considered yet. As we want to consider the phase shift due to a single input signal photon we have the normalization condition $\int_{T_1}^{T_2}dt' \left( {\hat {\mathcal E_{in}}}{\hat {\mathcal E_{in}}}^{\dagger}+{\hat {\mathcal E_{in}}}^{\dagger}{\hat {\mathcal E_{in}}}\right) = I$. Phase shift per signal photon to an atomic medium with exactly $N_g$ atoms in the ground state can now be calculated by replacing ${\hat {\mathcal E}}$ in Eq. (\ref{Phi_eq}) using Eq. (\ref{E_t}),
\begin{equation}
\Phi=\frac{4g^2/\kappa\Delta}{1+(2N_gg^2/\kappa\Delta)^2}.
\end{equation}

The term in the numerator, $4g^2/\kappa\Delta$, is the familiar dynamical stark shift enhanced by the cavity with decay rate $\kappa$. However, the phase $\Phi$ also has a term in denominator in this case, (1+$(2N_gg^2/\kappa\Delta)^2$), which depends on the number of atoms in the ground state $N_g$. This term originated from the atom-cavity interaction. Note that in our protocol where a coherent probe pulse is stored in the atoms before the signal passes above them, $N_g$ is not a constant. So, the phase $\Phi$ depends on $N_g$ and hence number of photons stored, which is not a constant. Coherent states by definition are in superposition of different photon number states as $\sum c_{n}|n\rangle$. If we use a coherent state with average photon number $\langle n\rangle = N_p$, $c_n$ = exp(-$N_p$/2)$\frac{N_p^{n/2}}{\sqrt{n!}}$ while $|n\rangle$ denotes a {\it n-} photon Fock state. After the probe is absorbed in the atomic memory it will correspond to an atomic state of $\sum c_{n}|N - n\rangle|n\rangle$ (a spin-coherent state), where the first and second state correspond to number of atoms in the ground state $|g\rangle$ and spin-ground state $|s\rangle$. {\it N} denotes the total number of atoms participating in the atomic ensemble memory. Hence we can define this photon number specific phase shift based on the probe photon number
\begin{equation}\label{phi_init}
\phi_{n} = \frac{4g^2/\kappa\Delta}{1+(2(N - n)g^2/\kappa\Delta)^2}.
\end{equation}

The term in the denominator of the phase varies with the square of probe photon number. This phase shift dependence on the number of stored probe photons occurs due to the presence of the atoms in the cavity which effectively shifts the cavity resonance. Hence, the signal photon experiences a detuning from the cavity and only part of the signal can enter the cavity leading to less phase shift of the atoms. This is reminiscent of the single atom conditional phase shift in a cavity \citep{duan2004scalable}. Although, here we are dealing with many photon probe state and hence many atoms are contributing to shifting the cavity resonance according to the probe's photon number distribution. Note that although only parts of the signal enter the cavity this does not affect the signal efficiency or fidelity as we are using an one sided cavity. We discussed this issue in more detail in section \ref{fidelity}.

The phase shift dependence on the number of stored probe photons can be compensated partially by making the cavity detuned from the input pulse. This will cancel the detuning that was coming as an off-set. For this we should detune the cavity by an amount $\frac{2ig^2}{\Delta}\langle\sigma_{gg}\rangle$ ($\langle\sigma_{ee}\rangle = 0$ as all the excited state atoms are transferred to the spin ground state). However even if this is incorporated, some residual dependence will still be present as a coherent probe pulse will have finite probabilities for different photon number states (Fock states) resulting in different amounts of phase shift based on the number of stored probe photons. Therefore, all these different phase contributions given by different Fock state components of a stored probe pulse cannot be all compensated simultaneously by detuning the signal. This residual phase shift dependence on the number of stored probe photons, due to the finite spread of the stored probe pulse in photon number states, will be important for our analysis. So, we attempt to understand this by analyzing what happens to the probe (or the atomic state generated by absorption of the probe) once the signal field has given it the phase shifts.

For our coherent probe pulse with average photon number $\langle n\rangle = N_p$ and a total of \textit{N} atoms participating in the atomic ensemble memory, we have $\langle\sigma_{gg}\rangle$ =  $N - N_p$. This implies a necessary detuning of the signal from the cavity by an amount $\frac{2ig^2}{\Delta}(N - N_p)$. If the new input electric field is ${\hat {\mathcal E}_{in1}} = {\hat {\mathcal E}_{in}}e^{\frac{2ig^2t}{\Delta}(N - N_p)}$ and the corresponding new electric field in the cavity is given by $\hat{{\mathcal E_c}}$, where ${\hat {\mathcal E_c}} = {\hat {\mathcal E}}e^{\frac{2ig^2t}{\Delta}(N - N_p)}$, we will have

\begin{equation}\label{E_c}
{\dot {\hat {\mathcal E_c}}}(t)= -\kappa {\hat {\mathcal E_c}}(t)+\sqrt{2\kappa}{\hat {\mathcal E}_{in1}}(t)+\frac{2ig^2}{\Delta}({\hat \sigma}_{gg}-(N-N_p)){\hat {\mathcal E_c}}(t),
\end{equation}
and hence a modified phase shift of
\begin{eqnarray}
\phi_{n} = \frac{4g^2/\kappa\Delta}{1+(2(N - n - (N-N_p))g^2/\kappa\Delta)^2} \nonumber
\\
= \frac{4g^2/\kappa\Delta}{1+(2(n - N_p)g^2/\kappa\Delta)^2}
\label{phi_n},
\end{eqnarray}
for a component of the probe pulse with \textit{n} photons (i.e. in $|n\rangle$ state).

For the free space case in \citep{sinclair2016proposal}, $\phi_{n} $ was independent of \textit{n}  $\forall$ \textit{n}, say $\phi_{n} = \phi$. A coherent state given by $|\alpha\rangle = e^{-|\alpha|^2/2}\sum\frac{\alpha^n}{\sqrt{n}} |n\rangle $ will transform to $e^{-|\alpha|^2/2}\sum\frac{\alpha^n}{\sqrt{n}}e^{in\phi} |n\rangle = |\alpha e^{i\phi}\rangle $ under such a phase shift for all its number state components. So, it will just become a phase-shifted coherent state.

In the cavity case, instead $\phi_{n} $ depends on $n$. If the term in the denominator of $\phi_{n} $ - $(2(n - N_p)g^2/\kappa\Delta)^2$ is large (close to 1 or larger) then the coherent state does not have an exact phase shift anymore. Instead the coherent state gets a scattered phase shift as depicted in Fig. \ref{q_func}.

\subsection{Husimi Q representation}

\begin{figure}[htbp]
\centering
\includegraphics[width=0.5\textwidth]{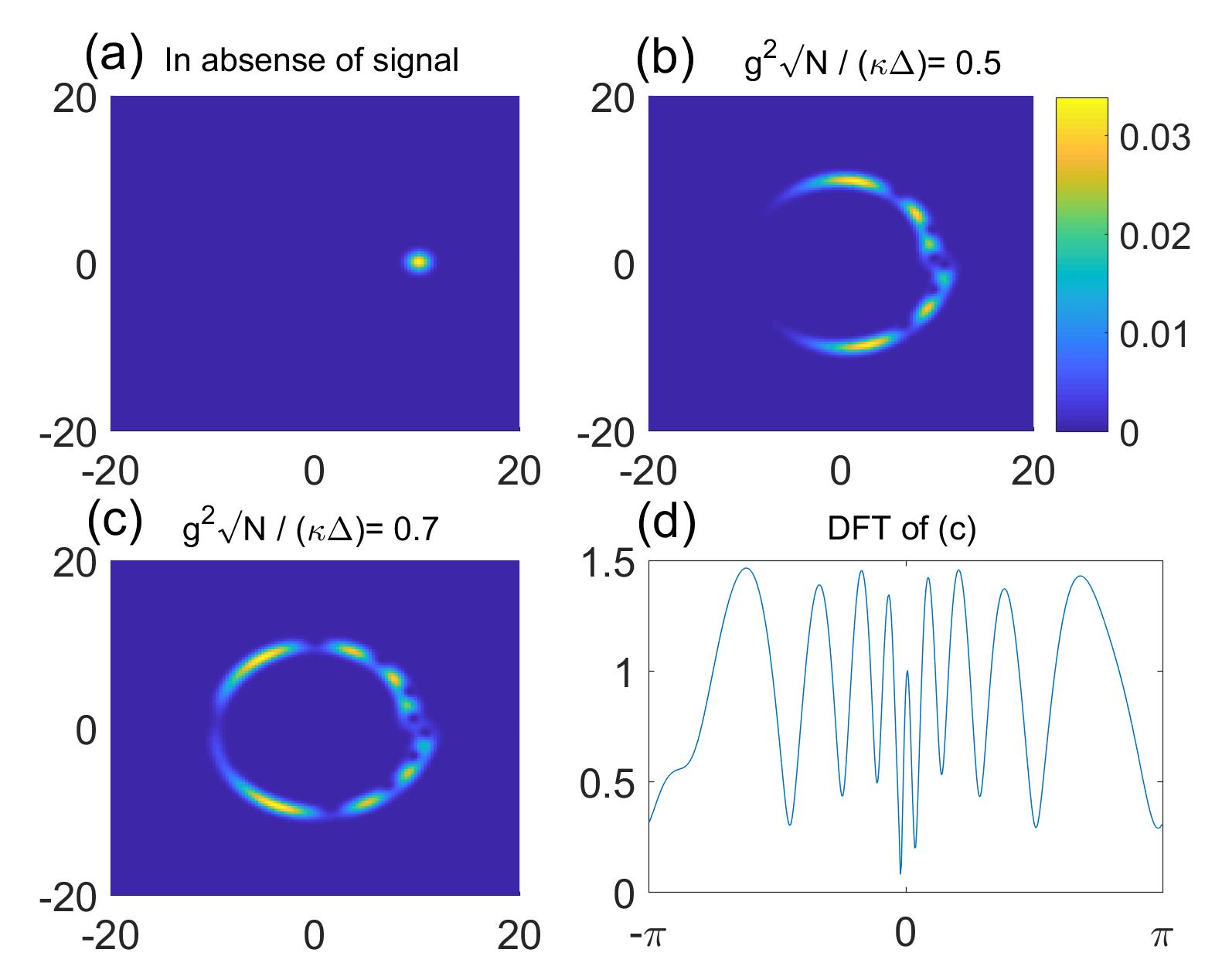}
\caption{(color online) (a)-(c) Color plot of the Husimi Q function in the phase space for the probe states with and without signal photon. The color map is shown beside (b). For all the plots the initial probe pulse is in a coherent state with average photon number $N_p$ = 100, and we assumed perfect storage and retrieval for simplicity. (a) Initial probe state. (b) Final probe state with parameters $g^2/\kappa\Delta = 0.5/\sqrt{N_p}$, showing that the state is slowly dispersing in phase. (c) Final probe state with parameters $g^2/\kappa\Delta = 0.7/\sqrt{N_p}$. Here the probe is completely dispersed in phase with very little probability left to be found in the location of the initial probe state. (d) A discrete Fourier transform(DFT) was done on probe state coefficients of (c). We plotted abs(DFT($c_n$)) to show the distribution over phase. The Fourier transform shows the exact same pattern in different phases that we already saw in the Q function.}
\label{q_func}
\end{figure}

The scattered nature of the phase shift is shown in Fig. \ref{q_func}, where a quasi-probability distribution of intial and final probe states are plotted in optical phase space using Husimi Q representation \citep{husimi1940some}.

In Husimi Q-representation, the quasi-probability distribution (or Q-function) of an optical state with density matrix $\hat{\rho}$ at a point $\alpha$ in phase space (corresponding to the center of coherent state $|\alpha\rangle$) is given by

\begin{equation}
Q(\alpha) = \frac{1}{\pi} \langle\alpha|\hat{\rho}|\alpha\rangle
\end{equation}

At a point in phase space, Q-function essentially calculates the overlap between the optical state and the coherent state centered on that point and hence is always positive. As we use pure states in our calculation for both the initial and  final probe state, we will write the density matrix  $\hat{\rho}$ = $|\psi\rangle\langle\psi|$ where $|\psi\rangle$ is the pure state.

In Fig.~\ref{q_func}(a)-(c), Q-function is plotted for different states in the optical phase space, so the  X and Y axes denotes the two conjugate optical quadratures. Fig.~\ref{q_func}(a) shows the Q-function of initial probe state is peaked at x =10, y =0 for a coherent state with average photon number $N_p$ = 100 and zero phase, i.e. $\alpha = 10$. In Fig.~\ref{q_func}(b) Q-function for the final probe, for parameter values $g^2/\kappa\Delta = 0.5/\sqrt{N_p}$, is plotted. Here, the probe state is somewhat scattered with contribution from positive and negative phases while maintaining the same photon number. This shows through a few oscillations of the Q-function at the same radius from center. However, it is still not completely dispersed in phase as for these parameters the noise term ($(2(n - N_p)g^2/\kappa\Delta)^2$) in the denominator of $\phi_n$ in Eq.~(\ref{phi_n}) is still not large enough. For $g^2/\kappa\Delta = 0.7/\sqrt{N_p}$ in Fig.~\ref{q_func}(c) the probe is completely dispersed in phase and there is negligible overlap with the initial probe state. Hence, we can in principle distinguish the initial and final probe state almost perfectly implying a successful QND measurement. This will be investigated in more details below. In (d) a discrete Fourier transform (DFT) of the probe state is carried out, i.e. of the number state($|n\rangle$) co-efficients $c_n e^{in\phi_n}$ with $c_n$ being initial coherent state co-efficients and $\phi_n$ given in Eq.~(\ref{phi_n}). As the photon number is maintained the fourier transform should indicate the variation of the probe state in quadrature phase. The same oscillatory structure in phase, exactly as in (c), are observed in (d).

\subsection{Inner product}\label{inner_pro_sec}

For successful non-destructive detection of the signal in our scheme, one needs to distinguish between the initial and final probe state practically with high probability. However, before considering practical protocols feasible for implementation, it needs to be ensured that these two states has negligible overlap. The overlap between the two states can be quantified by an inner product distance measure. This gives the minimum theoretical error probability in distinguishing the two states. Hence, the two states cannot be distinguished with a smaller error probability than $|$inner product$|^2$ using any protocol. This is a theoretical minimum.

\begin{figure}[htbp]
\centering
\includegraphics[width=0.5\textwidth]{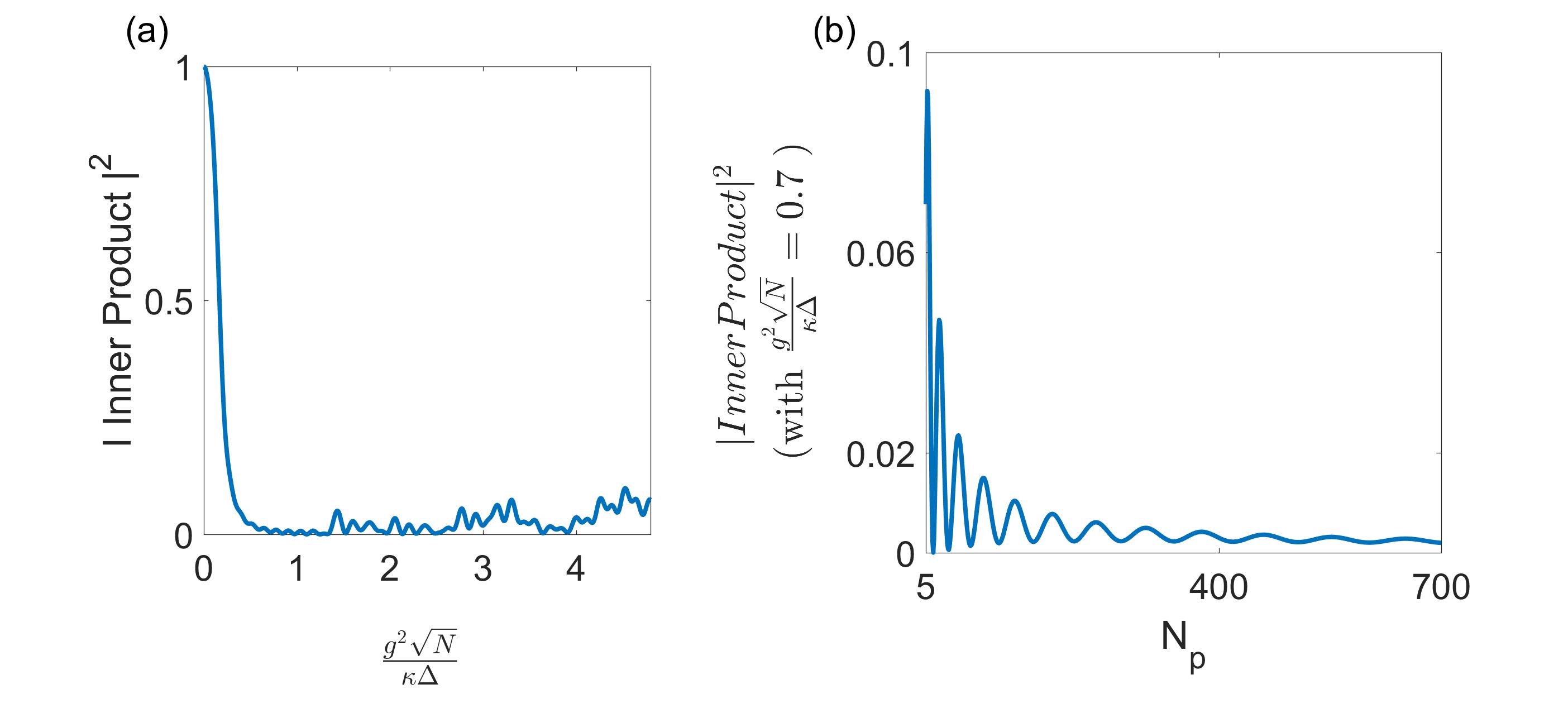}
\caption{Overlap between initial and final probe states, quantified as the square of the magnitude of inner product between these two states, (a) as a function of $g^2/\kappa\Delta$ for $N_p$ = 300 and (b) as a function of $N_p$ for $g^2/\kappa\Delta = 0.7/\sqrt{N_p}$. Subfigure (a) shows that in a certain parameter regime from $g^2/\kappa\Delta \sim 0.7/\sqrt{N_p}$ to $g^2/\kappa\Delta \sim 1.4/\sqrt{N_p}$ the overlap almost vanishes. Hence initial and final state can in principle be distinguished almost perfectly in this range. In (b) we see that the overlap decays almost as $1/\sqrt{N_p}$ (disregarding the oscillations) as explained in text. The characteristic oscillations of the system, introduced by the phase shift dependence on the number of stored probe photons as seen in Fig. \ref{q_func}, are visible in the overlap as well.}
\label{in_prs}
\end{figure}

The value of the $|$inner product$|^2$ changes as we change the value of $g^2\sqrt{N_p}/\kappa\Delta$. This is shown in Fig.~\ref{in_prs}~(a). The graph is plotted for values $N_p$ = 300. We see that around the parameter regime from $g^2/\kappa\Delta \sim 0.7/\sqrt{N_p}$ to $g^2/\kappa\Delta \sim 1.4/\sqrt{N_p}$ there is almost no overlap between the two states. Hence here the two states can in principle be distinguished perfectly. The graph shows a lot of fluctuations as $g^2/\kappa\Delta$ comes also in the denominator of our phase term as noise. Hence, to make our protocol robust against small experimental parameter fluctuations we should choose our $g^2/\kappa\Delta$ value so that it has minimal fluctuations, like places close to $g^2/\kappa\Delta = 0.7/\sqrt{N_p}$.

This value is used in Fig.~\ref{in_prs}~(b) to show $|$inner product$|^2$ variation with $N_p$. We can easily see that the overlap between initial and final probe states goes proportional to $1/\sqrt{N_p}$. The initial probe, being a coherent state will always have a spread of radius 1 in phase space. However the final probe state being spread all over the circle with photon number $N_p$ (considering it almost uniformly for simplicity), almost like a band of width 1, will correspond to a total length of $2\pi\sqrt{N_p}$. Therefore, the effective overlap between the two will decreases as $1/\sqrt{N_p}$. However, this is a very crude argument. There are oscillations in $|$inner product$|^2$ with $N_p$ (induced by the characteristic oscillations seen in Fig.~\ref{q_func}), but the overall trend scales as $1/\sqrt{N_p}$.

\subsection{Quadrature detection for practical discrimination}\label{quad_sub}

The initial and final probe state can be operationally discriminated through Homodyne detection. For that purpose the X quadrature is calculated for both of the states. In terms of photonic annihilation and creation operators $\hat{a}$ and $\hat{a}^{\dagger}$ as the X quadrature operator can be written as $\hat{X}$ = $\hat{a}$ + $\hat{a}^{\dagger}$. If eigenstates of $\hat{X}$ are represented as $|x\rangle$ then we know from the study of simple harmonic oscillators

\begin{equation}
|x\rangle = \sum_n \frac{H_n(x)}{(2^n n!)^{1/2}} \frac{e^{-x^2/2}}{\pi^{1/4}}|n\rangle
\label{quad_x}
\end{equation}
where $H_n(x)$ is a $n^{th}$-order Hermite polynomial evaluated at point $x$. To calculate the quadrature measurement probability density at a quadrature value of $x$ for our probe in a particular quantum state $|\psi\rangle$ we calculated the value of $|\langle\psi|x\rangle|^2$.

\begin{figure}[htbp]
\centering
\includegraphics[width=0.5\textwidth]{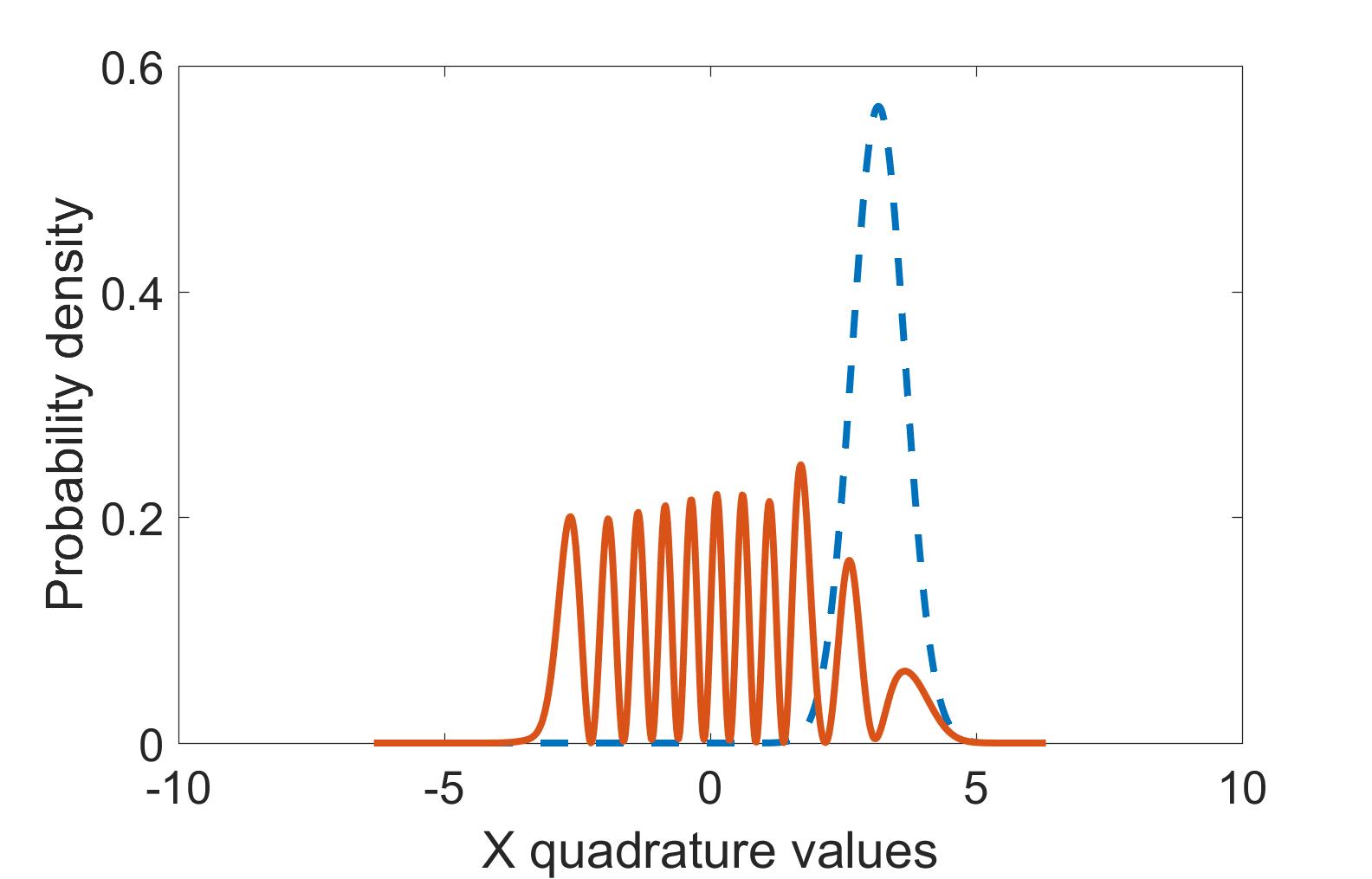}
\caption{X Quadrature measurement probability density of probe states with (red solid line) and without (blue dashed line) a signal photon present for probe photon number $N_p$ = 10 and $g^2/\kappa\Delta = 0.7/\sqrt{N_p}$.}
\label{quadr_N_10}
\end{figure}

For $g^2/\kappa\Delta = 0.7/\sqrt{N_p}$ and $N_p$ = 10 the X quadrature detection probabilities are shown in Fig \ref{quadr_N_10}. Now, let us make a cut off (say, 1.64 in the above case) so that a the initial probe state has its X quadrature value higher than 1.64 with a high probability (99.9$\%$ in the above case). Now, whenever we get a X quadrature measurement of our probe below that we decide in favor of the final probe state and say that a single photon passed through in those cases. In this way, we will only be able to detect a signal with a certain success probability (e.g. 72$\%$ for $N_p$ =10 as shown in Fig \ref{quadr_N_10}). However the probability that we make a false positive decision about the presence of signal while it is not there is very low, only 0.1$\%$.

If we allow 1$\%$ or 10$\%$ error rate in detecting false positives (which will correspond to initial probe state probabilities above the cut-off to be 99$\%$ and 90$\%$ respectively) we will get success probabilities of detections of 81$\%$ and 85$\%$ respectively.

The success rates corresponding to different probe photon numbers are presented in Table \ref{quad_table} for false positive detection (i.e. deciding there was a photon when there was none) probabilities of 0.1 $\%$ and 1 $\%$.

\begin{table}[htbp]
\begin{tabular}{ | p{2.5cm} | p{3cm} | p{3cm} |}
    \hline
    Probe photon Number ($N_p$) & Success rate of detection for 0.1$\%$ error rate  & Success rate of detection for 1$\%$ error rate \\ \hline
    10 & 72.29$\%$ & 81.46$\%$ \\ \hline
    30 & 76.27$\%$ & 83.62$\%$ \\ \hline
    50 & 80.80$\%$ & 86.31$\%$ \\ \hline
\end{tabular}
\caption{The success rate of our single photon QND proposal is shown for different values of probe photon number ($N_p$) and for $g^2/\kappa\Delta = 0.7/\sqrt{N_p}$. The success rates are mentioned for two different false positive detection probabilities of 0.1 $\%$ and 1 $\%$.}
\label{quad_table}
\end{table}

One thing to be noted here is that the $|$inner product$|^2$ between the probe states with and without a signal photon for $N_p$ =10 was 0.0964. So, any type of measurement on the system with vanishing error rate can only give you a maximum of 90.36$\%$ success rate. For the small error rate of 0.1$\%$ we got 72.29$\%$ success rate for $N_p$ = 10. For vanishing error rates, we couldn't get very close to the theoretical maximum as the X quadrature measurement was not providing the perfect discrimination, at least for small $N_p$.

The different values between the theoretical discrimination by inner product and in a practical approach like quadrature measurement originates from the fact that in a practical approach we always measure probability distribution (e.g. $|\langle\psi|x\rangle|^2$ for quadrature) and not the probability amplitude (i.e. $\langle\psi|x\rangle$). Hence, even if two states have vanishing inner product they can not be distinguished by quadrature detection completely as long as they overlap in quadrature values. However, this may not be an insurmountable difficulty. Other clever measurements schemes (e.g. invoking interference) may be constructed such that the states do not overlap in the corresponding observable. In such an ideal measurement scheme crafted for two particular states the practically achievable discrimination should approach the theoretical minimum predicted by inner product.

On a more practical consideration, the values in Table \ref{quad_table} shows that success rate for the quadrature detection increases with increasing $N$. This is expected as the overlap, i.e. $|$inner product$|^2$, between the initial and final probe states decrease with increasing $N$ as shown in Fig. \ref{in_prs}(b) (except for the small characteristic oscillations). This is due to the fact that the final probe state gets spread over a larger radius in the phase space while the spread in the initial probe state remains constant in the phase space, as discussed in Section \ref{inner_pro_sec}. So, for a much larger $N$ we would be able to nearly perfectly distinguish the probe states in presence and absence of a signal photon even using the practical method of quadrature detection.

\subsection{Signal Fidelity for Time-Bin Qubit Detection}\label{fidelity}

In this section, we analyze the signal fidelity of the output signal from the cavity once the signal has imparted the necessary phase shift. As we discussed after Eq. (\ref{phi_init}), due to the phase shift dependence on the number of stored probe photons some parts of the signal gets detuned from cavity and does not impart phase shift to the atoms for all $|n\rangle$, where $n$ is a particular Fock state component of the probe pulse. This remains even after phase compensation, performed by detuning the input signal, due to the residual phase shift. The detuned portion of the signal gets reflected and do not contribute to the phase shift. However, this does not reduce signal fidelity or efficiency as we use a single-sided cavity with a fast decay rate compared to signal bandwidth. As a single-sided cavity is used the input signal reflected from front and the back mirror interferes and due to the fast cavity decay rate there is almost no time lag to form the output signal. We show this mathematically in the following. Using the input-output relation \citep{scully1999quantum} for a one sided cavity, ${\hat {\mathcal E}}_{out}(t)=\sqrt{2\kappa}{\hat {\mathcal E_c}}(t)-{\hat {\mathcal E}}_{in}(t)$, we find from Eq.~(\ref{E_c})

\begin{equation}
{\hat {\mathcal E}}_{out}(t)=\frac{\kappa+\frac{2i(N_p -n) g^2}{\Delta}}{\kappa-\frac{2i(N_p-n) g^2}{\Delta}}{\hat {\mathcal E}}_{in}(t).
\end{equation}

So, the output signal field differs from the input field by only a global phase($\theta$) of magnitude 2$\times$arg($\kappa+\frac{2i(N_p-n) g^2}{\Delta}$) or $\theta$ = 2 tan$^{-1}(\frac{2(N_p-n) g^2}{\kappa\Delta})$. This implies $|{\hat {\mathcal E}}_{out}(t)|^2 = |{\hat {\mathcal E}}_{in}(t)|^2$ for all values of $n$.

If the signal photon is in a time-bin qubit then we need to maintain coherence between the early and late time bins. Note, although the ${\mathcal E}_{out}$ gets different phases for different values of $n$, this does not affect time-bin qubit state as both the early and late time bins pass over the same atomic ensemble at a small time difference. If the time lag between early and late qubit is $T$, the coherence needs to be maintained within this time interval implying $(N_p - n)$ needs to remain constant in that time interval. Here, $n$ denotes a specific  photon number state which is absorbed in the atoms and the corresponding atomic excitation are transferred to another ground state. Hence, $n$ correspond to excitations in a second ground state, which may decay with time. If the rate of decay of the second ground state is $\gamma_s$ then with time expectation value of $n$ turns into $n e^{-\gamma_s T}$. If $\gamma_s T << 1$, then the change in $n$ goes as $\Delta n = n(1-e^{-\gamma_s T}) \approx n \gamma_s T$. Now, for small changes in $n$ we can represent the change in magnitude of $\theta_n$ as $\Delta \theta_n$ where,

\begin{equation}
\Delta \theta_n = \frac{\frac{2g^2}{\kappa\Delta}}{1+(\frac{2(N_p -n)g^2}{\kappa\Delta})^2} \Delta n =  \frac{\frac{2g^2}{\kappa\Delta}}{1+(\frac{2(N_p -n)g^2}{\kappa\Delta})^2} n \gamma_s T.
\end{equation}

We include the the initial and final probe state discrimination condition of $\frac{g^2}{\kappa\Delta} \approx \frac{1}{\sqrt{N_p}}$ obtained from inner product analysis. The maximum value of $\Delta \theta_n$ is around $n = N_p$ or $(\Delta \theta)_{max} \approx 2 \sqrt{N_p}\gamma_s T $ for $\gamma_s T << 1$. For $\Delta \theta = \pi$ the phase between two time bins flip and as the phases will be different for different values of $n$ this will severely limit the signal fidelity. Hence we will need to have $(\Delta \theta)_{max} << \pi$ for high fidelity signal output.
Later in Section~\ref{Implement_sec}, we estimate  $N_p$ = 6000. So for a signal bandwidth of 1 MHz (i.e $T$ = 1 $\mu$s) and a moderate spin ground state dephasing rate $\gamma_s = 0.34 kHz$ at 5K temperature in Nd \citep{wolfowicz2015coherent}, we have $(\Delta \theta)_{max} \approx 0.0527 \approx \pi/60$.

Given the probe state and hence the stored atomic excitations are in a coherence state with coefficients of photon number state $|n\rangle$ given by $c_n$ = exp(-$N_p$/2)$\frac{N_p^{n/2}}{\sqrt{n!}}$, we have signal fidelity for time-bin qubit as $\sqrt{\sum_n |c_n|^2|\frac{1+e^{i\theta_n}}{2}|^2}$. For the above mentioned parameters, where we store the probe pulse in a second spin ground state with a long lifetime signal fidelity is 0.9999. Instead if the probe is stored in the excited state using a different protocol, e.g. atomic frequency comb quamtum memory protocol \citep{afzelius2009multimode}, we will have the decoherence rate as $\gamma_h = 100 KHz$ for our doping as mentioned in Section \ref{Implement_sec}. In that case, with a 1 MHz bandwidth signal the fidelity drops to 0.6915 but using a 10 MHz bandwidth signal we will acquire a fidelity of 0.9216.

\section{Signal loss}\label{loss_sec}

In order to analyze off-resonant absorption loss for a cavity-enhanced signal we use the total Hamiltonian in Eq.~(\ref{H_tot}).
This results in
\begin{multline}\label{sigma_cavity}
{\dot{\hat \sigma}}_{eg}(t;\delta)=(-\gamma+i\delta){\hat \sigma}_{eg}(t;\delta)\\
-ig{\hat {\mathcal E^{\dagger}}}e^{-i\Delta t}({\hat \sigma}_{gg}(t;\delta)-{\hat \sigma}_{ee}(t;\delta))
\end{multline}
The dynamics of the cavity field and the cavity input-output relation is given by
\begin{align}\label{cavity_field}
&{\dot {\hat {\mathcal E}}}(t)=-\kappa {\hat {\mathcal E}}+\sqrt{2\kappa}{\hat {\mathcal E}}_{in}(t)+ige^{-i\Delta t}\sum_{\delta}{N(\delta){\hat \sigma}_{ge}(t;\delta)}\\
&{\hat {\mathcal E}}_{out}(t)=\sqrt{2\kappa}{\hat {\mathcal E}}(t)-{\hat {\mathcal E}}_{in}(t).
\end{align}
Given that the single excitation wavefunctions are governed by the same equations, we can find the steady state solution to these equations by taking the Fourier transform of Eqs.~(\ref{sigma_cavity}) and (\ref{cavity_field}). Taking the Fourier transform of Eq.~(\ref{sigma_cavity}) gives,
\begin{equation}\label{sigma_cavity_sol}
{\tilde \sigma}_{eg}(\omega;\delta)=\frac{-ig}{i(\omega-\delta)+\gamma}{\tilde {\mathcal E}}^*(\omega-\Delta).
\end{equation}
Using this result, and assuming that $\Delta\gg\delta$ $\forall \delta$ i.e. $\Delta$ is larger than the inhomogeneous linewidth of the atoms considered, we can simplify the resulting expression for the cavity field to
\begin{equation}\label{cavity_field_sol}
{\tilde {\mathcal E}}(\omega)=\frac{\sqrt{2\kappa}}{i\omega+\kappa+\frac{ig^2N}{\omega-\Delta+i\gamma}}{\tilde {\mathcal E}}_{in}(\omega).
\end{equation}
Using this result and the cavity input-output relation we can find the cavity output field. For the case where the signal bandwidth is smaller than the signal-atom detuning $\Delta$, we can assume that the loss will be uniform and therefore analyze the cavity output field at $\omega=0$. This is given by,
\begin{equation}\label{cavity_output}
{\tilde {\mathcal E}}_{out}(\omega=0)=\left( \frac{2\kappa}{\kappa-\frac{ig^2N}{\Delta-i\gamma}}-1\right) {\tilde {\mathcal E}}_{in}(0).
\end{equation}
In order to estimate the loss, we find the output intensity with respect to the input field intensity.
\begin{eqnarray}\label{cavity_loss}
&|{\tilde {\mathcal E}}_{out}(0)|^2=\alpha  |{\tilde {\mathcal E}}_{in}(0)|^2 \\ \nonumber
&\alpha=\left( 1-\frac{4\gamma g^2N}{\kappa\Delta^2}+(\frac{2\gamma g^2N}{\kappa\Delta^2})^2+(\frac{2 g^2N}{\kappa\Delta})^2+{\mathcal O}(1/\kappa^3) \right).
\end{eqnarray}
Given that we assume $\kappa$ to be the fastest rate in the system the main contribution to loss is given by
\begin{equation}\label{cavity_loss_formula}
\zeta=\frac{4\gamma g^2N}{\kappa\Delta^2}.
\end{equation}

However, the atoms are also within the cavity. If they are not completely off-resonant, spontaneous emission is enhanced by the Purcell factor $ \frac{3Q}{4\pi^2}\frac{(\lambda_0/n)^3}{V}\frac{(\kappa/2)^2}{(\kappa/2)^2+\Delta^2}$. Considering this possible enhancement effect on spontaneous emission, the formula for the signal loss in cavity, i.e. Eq. (\ref{cavity_loss_formula}), becomes - $\frac{4\gamma_r g^2 N}{\kappa\Delta^2} \frac{3Q}{4\pi^2}\frac{(\lambda_0/n)^3}{V}\frac{(\kappa/2)^2}{(\kappa/2)^2+\Delta^2}$. Note that now homogenous linewidth, $\gamma$ is replaced by radiative linewidth $\gamma_r$. This is because cavity enhances the radiative linewidth $\gamma_r$ to $\gamma_r \frac{3Q}{4\pi^2}\frac{(\lambda_0/n)^3}{V}\frac{(\kappa/2)^2}{(\kappa/2)^2+\Delta^2}$ and in case of a large enhancement that becomes the major contributing factor in the homogenous linewidth and hence in the resulting dephasing.

In the phase shift analysis we saw that $\frac{g^2}{\kappa\Delta} \sim \frac{1}{\sqrt{\eta_r N}}$ entails faithful discrimination of the probe pulse, where $\eta_r$ is the probe retrieval efficiency and \textit{N} is the number of atoms excited by the probe. Based on the definition of spontaneous emission rate in a solid where dipoles are oriented in one direction $\gamma_r =\frac{\mu_{eg}^2 k_s^3}{\pi \epsilon_0 \hbar}$ and single photon coupling $g=\mu_{eg}\sqrt{\frac{\omega_s}{2\hbar\epsilon_0 V}}$, we find that $\frac{g^2}{\kappa\Delta} = \frac{1}{4\pi}\frac{\lambda_0^2 }{n^2 A}\frac{F\gamma_r}{\Delta} $. Here $F$ is the finesse of the cavity and it is related to the cavity quality factor as Q = $F\frac{2L}{\lambda_0/n}$. Combining these formulas we find the cavity enhanced loss

\begin{equation}
\frac{4\gamma_r g^2 N}{\kappa\Delta^2} \frac{3Q}{4\pi^2}\frac{(\lambda_0/n)^3}{V}\frac{(\kappa/2)^2}{(\kappa/2)^2+\Delta^2} = \frac{6}{\pi \eta_r}\frac{(\kappa/2)^2}{(\kappa/2)^2+\Delta^2}.
\end{equation}

Considering ideal retrieval $\eta_r \sim 1$, we find the expression for loss to be - $\frac{2}{\eta_r}\frac{(\kappa/2)^2}{(\kappa/2)^2+\Delta^2}$. Hence the only way to have low loss is having a high value of $\Delta$ compared to $\kappa/2$. For $\Delta = 3 (\kappa/2)$ we have around $20\%$ loss, while $\Delta = 3\kappa$ amounts to only $5.4\%$ loss. Until now we have considered the ideal case with $\eta_r = 1$, but for practical purposes that may not be achievable. However an $\eta_r = 0.4$ may well be achievable as that amounts to a total memory efficiency of 16$\%$ only (considering storage and retrieval efficiency to be identical). That will not make a huge change in the corresponding values of $\Delta$ for similar loss probabilities. $\Delta$ values will only need to be multiplied by a factor of approximately $\sqrt(1/0.4) \sim 1.6$, i.e. $\Delta = 2.4 \kappa$ and $\Delta = 4.8 \kappa$ for 20$\%$ and 5.4$\%$ loss rates respectively when $\eta_r = 0.4$.

\section{Implementation}\label{Implement_sec}

Implementation of the proposal in rare earth ion-doped crystals has three stages - storing the probe, imparting a significant phase shift by single photon level signal and finally measuring the retrieved probe to know the presence of the signal. We already discussed about the probe storage in Section \ref{probe_sec}. There, we mentioned the necessity of dynamically detuning the cavity to facilitate storage. We will return to this in more detail later in this section. Almost all of this section is dedicated to the next stage which is estimating the imparted phase shift to the probe. This is because much of the requirements for a proposed system are decided based on this stage. In the final stage, the probe pulse needs to be measured to distinguish between the probe states with and without the signal. In Section \ref{quad_sub} we suggested to perform a quadrature detection for this purpose by means of homodyne detection, which is a standard optical measurement scheme.

\subsection{Proposed parameter regime}

The principal requirements for implementation of our proposed scheme are dictated by the ability to impart a large enough phase shift. Hence, we shall first return to the phase shift requirements for our proposal. The inner product analysis in Section \ref{inner_pro_sec} of our theoretical model shows $g^2/\kappa\Delta \approx 1/\sqrt{\eta_r N_p}$ needs to be satisfied to distinguish between the probe states with and without a signal photon. This can be re-arranged to write it in terms of the factor $f = \frac{g^2}{\kappa\Delta} \sqrt{\eta_r N_p} \approx 1$. By considering single photon coupling  $g=\mu_{eg}\sqrt{\frac{\omega_s}{2\hbar\epsilon_0 V}}$ and radiative transition rate $\gamma_r=\frac{\mu_{eg}^2 k_s^3}{3\pi \epsilon_0 \hbar}$ this condition is equivalent to $f = \frac{1}{4\pi}\frac{\lambda_0^2}{n^2A} \frac{F \gamma_r}{ \Delta} \sqrt{\eta_r N_p} \approx 1$, where $n$ is the refractive index inside cavity and $F$ is the finesse. Considering $N$ atoms inside the cavity mode-volume $V$, we have $N_p \propto N \propto V$. Hence, we conclude that $f \propto  \frac{ \gamma_r}{ \Delta}F\sqrt{\frac{L}{A}}$; note that it depends linearly on the finesse, but only on the square root of the length.

This analysis shows that the implementation of the proposal in rare earth ion doped ensemble demands a high finesse, small transverse area and preferably long cavity. Nanophotonic rare-earth ion coupled cavities are being fabricated in photonic crystal cavities etched inside rare-earth ion doped crystals \cite{zhong2015nanophotonic, zhong2016high, zhong2017interfacing, faraon2018controlling}, in silicon photonic crystal cavity evanescently coupled to rare-earth ions \citep{jeff2017isolating} or in fiber tip microcavities containg rare-earth ion doped nanocrystals \citep{hunger2018cavity}.

The rare-earth ion, which will be doped in such a cavity to interact with the photon, will require a large dipole moment for higher atom-photon coupling (higher g and so $\gamma_r$) to increase the phase shift. For our estimates we have chosen neodymium (Nd$^3+$ in Nd:YVO) as it is one of the rare earth elements with a higher dipole and high optical coherence time \citep{afzelius2010efficient, hastings2008spectral}. Optical coherence time is important as we are going to use AFC quantum memory protocol for probe storage in the excited state. We shall be using the Z$_1$ to Y$_1$ levels in Nd$^{3+}$:$^4I_{9/2} \rightarrow  {}^4F_{3/2}$ transition at 879 nm (see Fig. 1 of \citep{afzelius2010efficient}). This levels in Nd:YVO are particularly useful as each of these Z$_1$ and Y$_1$ levels (Krammers doublet) split into two levels creating a four level system with favorable selection rules under an applied magnetic field along the YVO crystal axis \citep{afzelius2010efficient}. The selection rules are such that light polarized along the crystal axis (or perpendicular to it) interacts only with each set of sub-levels and there is no cross-talk between them (or vice-versa). This effectively creates convenient $\Lambda$ systems inside the four level system. In \citep{afzelius2010efficient} it is experimentally shown that the branching ratio between the direct and cross-transitions is 95$\%$-5$\%$, which is quite close to a perfect selection rule. In our proposal both the probe and signal will be polarized along the crystal axis and interact with only one sub-level, as both light and the cavity will be far detuned from the other sub-level. The two sub levels will be far detuned by a large applied magnetic field. Although, both signal and probe only interact with one sub-level we will still use the $\Lambda$ system for optical pumping to prepare the AFC quantum memory for probe storage.

The only experimentally free parameter in the phase shift formula (and hence in $f$) is $\Delta$ which can be decreased to increase the phase shift. However, $\Delta$ is constrained by signal loss. As shown in Section \ref{loss_sec}, signal photon loss on resonance with a cavity with high quality factor is given by $\frac{2}{\eta_r}\frac{(\kappa/2)^2}{(\kappa/2)^2+\Delta^2}$ where the cavity enhanced spontaneous emission dominates the decoherence process. Hence, a large detuning compared to cavity linewidth (around $\Delta > 3\kappa $) is necessary for low loss.

We are now using the AFC storage protocol in Nd:YVO photonic crystal cavities \citep{zhong2017nanophotonic} as an example to provide an estimate for implementation of the scheme. The main condition for successful implementation is to reach the phase shift condition $g^2/\kappa\Delta \sim 1/\sqrt{\eta_r N_p}$ while simultaneously having $\Delta \geq 3\kappa$ to keep the loss low. Here, we propose one set of parameters to reach the desired regime - $g = 2\pi \times$ 8 MHz, $\kappa$ = 2$\pi \times$ 30 MHz, $\Delta$ = 2$\pi \times$ 100 MHz, $\eta_r$ = 0.5 and $N_p$ = 6000. This yields $f = \frac{g^2}{\kappa\Delta} \sqrt{\eta_r N_p}$ = 1.16 which is around 1 and hence sufficient for a successful QND detection of a single photon. The corresponding value for loss is $\frac{2}{\eta_r}\frac{(\kappa/2)^2}{(\kappa/2)^2+\Delta^2} = 0.073$ or 7.3$\%$. For $N_p$ = 6000 the probability to distinguish between the probe states with and without a signal photon present by an $X$ quadrature measurement is very high. The probe state that interacted with signal is scattered all over a circle in phase space with a radius of around 77 ($\sim \sqrt{6000}$ ), while the probe state without the signal is a coherent state, which is highly localized. Based on this, we estimate the probability of overlap between the $X$ quadrature distributions of the two states to be less than 4$\%$. Hence, we can distinguish between the two states with very low error rates with a success probability of over 96$\%$. Incorporating the effect of the 7.3$\%$ loss this would produce around 89$\%$ success probability in total.

Nanophotonic cavities built in Nd:YVO have already achieved experimental quality factors around 20,000 \citep{zhong2016high} which corresponds to a cavity linewidth $\kappa = 2 \pi \times 17$ GHz. Achieving $\kappa=2 \pi \times 30$ MHz will probably require a combination of increasing the finesse and the length of the cavity. This may be realistic given the steady and fast recent progress in building high quality factor photonic crystal cavities  \citep{faraon2018controlling, jeff2017isolating, hunger2018cavity}. For the Nd:YVO system Ref. \citep{zhong2016high} suggests that it may be possible to improve the finesse by an order of magnitude or more by changes in the fabrication process such as decreasing the sidewall angle for the nanocavities and post-fabrication annealing. Increasing the length of the cavity should also be possible, but will require longer milling times (for ion-beam based fabrication). Having sufficiently many ions in the cavity to be able to store $N_p=6000$ photons as suggested above probably requires an increase in the cavity length by at least an order of magnitude, taking into account the fact that the AFC memory protocol requires spectral tailoring, which reduces the available number of atoms. Another attractive way to increase the number of atoms would be to use recently developed stoichiometric rare-earth crystals \citep{ahlefeldt2013precision,ahlefeldt2016ultranarrow} where ultranarrow inhomogenous linewidth has been observed. However, currently these crystals are made only from weak dipole elements like Eu$^{3+}$ which is not good for our proposal. A nanocavity etched in a stoichiometric crystal, made of an rare-earth element with a strong dipole element, would definitely be useful as many more atoms can be accommodated inside the cavity.

The other approach towards ensemble QND measurements can be increasing atom-cavity coupling or $g$ value \citep{faraon2018controlling}. One way towards this is by decreasing cavity mode volume through incorporation of dielectric discontinuities \citep{englund2017self} into cavity design. However, the number of available atoms for phase shift also decrease with decreasing cavity mode volume as reflected in $f \propto F\sqrt{\frac{L}{A}}$. So, for the purpose of ensemble QND only decreasing the cavity transverse area will help while decreasing cavity length to decrease mode volume will adversely affect the ensemble QND detection. Another strategy to increase the coupling factor may be to change the AC stark shift interaction to higher dipole 4f $\leftrightarrow$ 5d transition while storing the probe using the 4f $\leftrightarrow$ 4f transition which has desireable optical and spin coherence properties. However, this will need a doubly resonant cavity \citep{faraon2018controlling}.

Recently, there has also been development of other attractive nanocavity systems with rare-earth ions incorporated into them \citep{jeff2017isolating, hunger2018cavity}. In \citep{jeff2017isolating} a Si-photonic crystal cavity was manufactured through which light is evanescently coupled to a single Er$^{3+}$ ion, present inside a lightly doped Er:YSO crystal. This is an attractive system with cavity quality factor of 51,000 for the Er$^{3+}$ transition at 1.5 $\mu m$ wavelength. This led to the coupling of individual Er$^{3+}$ ions to the cavity. Similar system can probably be constructed for higher dipole moment rare earth ions like Nd$^{3+}$ but the rare earth ions are evanescently coupled here which decreases the cavity coupling g and Purcell factor compared to what will have been possible if they were present inside the cavity. The evanescent coupling also decreases the number of atoms that can be coupled to the cavity.

Another system containing a Eu$^{3+}$ doped nanocrystal inside a free-space cavity between a fiber tip and a mirror is introduced recently \citep{hunger2018cavity}. This also have attractive cavity parameters of finesse 17,000, $\kappa$ = 1.3 GHz and a corresponding cavity quality factor of around 400,000. The much higher finesse and quality factor of the cavity paves the way for implementing our proposal. However, this design uses a nanocrystal of dimensions 40-60 nm. Hence, not a lot ions can be accommodated inside the crystal, at least inside a reasonable frequency range of 100 MHz-2 GHz which may impose some limitations. A short frequency range is required so that the detuning ($\Delta$) doesn't need to be too large.

\subsection{Dynamical switching of the cavity}

	After the probe storage we need to detune the cavity by $\Delta \sim$ 100 MHz for it to be on resonance with the signal and later detune it back to retrieve the probe. The cavity is initially in resonance with the atoms at $\nu \sim$ 340 THz frequency. If the original cavity length is \textit{L} and the change in length needed to detune the cavity by an amount $\Delta$ (in frequncy) is $\Delta L$ then $\frac{\Delta L}{L} = \frac{\Delta}{\nu} \sim 3\times 10^{-7}$. Considering the above strain and the Young's modulus of the YVO crystal = 133 GPa \citep{peng2001study}, we can calculate the necessary stress to be 44 kPa. Here, we are taking a commercial piezo detector as an example, P-882.1 in \citep{WinNT}. This has a 6 mm$^2$ surface area. Hence, 44 kPa stress will correspond to a applied force of 0.26 N. Piezo can in general deliver far higher magnitude maximum  forces, denoted as block forces than this \citep{yang2004leak, WinNT}. In this specific case of \citep{WinNT}, the piezo actuator has a block force of 190 N corresponding to a maximum displacement of 8 $\mu$m . Hence the force necessary in our case is only about 0.14 $\%$ of the maximum force. For applying such a small force resolution becomes important. Piezos also generally have sub-nanometer resolution in precise positioning \citep{WinNT}. For the maximum displacement of 8 $\mu$m  block force is 190 N \citep{WinNT} and for sub-nanometer resolution the minimum force produced will be at most 0.023 N. We need a force of 0.26 N hence this gives us at most a 10.8 $\%$ error or about 11 MHz error in positioning the 30 MHz cavity. This can affect our phase shift to some amount. This commercial piezo actuator \citep{WinNT} has microsecond response times which is around the AFC storage times. Also there has been a lot of ongoing research on sub-microsecond piezo \citep{judy2009piezoelectric}.

The other issue with the strain given by piezo is that this deforms the crystal structure which causes a small stark shift between Nd levels. To the best of our knowledge, there is no experimental data in the literature on the strain (or equivalently stress) induced stark shift of Nd:YVO. However, experiments have been performed on other crystals like Nd:YAlO$_3$ \citep{hua1997pressure}. In Nd:YAlO$_3$ the magnitude of stress induced shift in Z$_1$ to Y$_1$ transition in Nd$^{3+}$:$^4I_{9/2} \rightarrow  {}^4F_{3/2}$ levels is found to be around 32.05 Hz/Pa (using Eq. (2b) in \citep{hua1997pressure}). This will imply 1.41 MHz of stress induced shift to our desired transition. This is small compared to our required detuning of 100 MHz. For an approximately linear rise of the piezo actuator over 50 ns time this will give a phase shift to the probe of magnitude about 2$\pi \times$0.035 rad. But, this shift will not affect our proposal adversely as this phase shift will be present independent of the presence of the signal. However, an unpredictable error in the piezo displacement will affect the proposal. This is because the probe will be unpredictably phase shifted and hence on making a quadrature measurement there will be some probability for false positive result. So, for a $5\%$ error in piezo displacement this will result in a phase shift of 2$\pi \times$0.035. Our proposal has a similar quantity like phase shift $ g^2/\kappa\Delta \sim 1/\sqrt{6000}$ = 0.012, and hence naively, there should be about 30$\%$ error. But, the probe will not be scattered over the phase space due to this phase shift as it occurs due to the Stark shift in the atoms and has nothing to do with the presence of the cavity. Hence, the probability of false positive detection will be very low even for a phase shift comparable to the 2$\pi \times$ 0.012 rad value as due to the cavity in presence of a signal final probe state quadrature gets scattered all over the phase space, i.e. in a 2$\pi$ rad angle. So, this strain induced stark shift will only decrease the success probability by a very small amount. If we consider that the final probe states spreads uniformly this will correspond to roughly 0.012/1 $\approx 1 \%$ less success probability.

Another attractive approach toward achieving the dynamical detuning is to detune the atoms, instead of the cavity, by Zeeman effect using an external magnetic field. A magnetic field is already present to enforce the selection rules. The magnitude of this magnetic field needs to be increased to give an extra 100 MHz detuning between the atoms quickly. Similar to the piezo strain induced shift, in the Zeeman shift process the dynamical fluctuations while detuning the atoms will affect the phase shift.

\subsection{Combining Cavity and Multipass Approach}

Another avenue towards implementation can be combining the cavity approach with the multipass approach described in \citep{sinclair2016proposal}. In the multipass approach instead of using a cavity the signal is simply passed multiple times through the crystal (in a waveguide) using an optical switch. However, multipass arrangement alone can not achieve single photon QND as it will require a lot of passes ($>$ 200) \citep{sinclair2016proposal} which is currently not feasible as the optical switches, essential for the multipass arrangement, cause far too much switching loss for so many passes. However, if we combine the multipass arrangement with the cavity then we may be able to restrict ourselves with much fewer passes. The principle advantage of this scheme is that loss can be decreased significantly without having a large detuning, which will improve the phase shift considerably.

In the multipass arrangement as the photon passes over the atoms multiple(say \textit{m}) times both the phase shift and loss gets multiplied by \textit{m} \citep{sinclair2016proposal}. i.e. phase = $m\frac{g^2}{\kappa\Delta} \sim \frac{1}{\sqrt{\eta_r N}}$ and loss = $m\frac{4\gamma_r g^2 N}{\kappa\Delta^2} \frac{3Q}{4\pi^2}\frac{(\lambda_0/n)^3}{V}\frac{(\kappa/2)^2}{(\kappa/2)^2+\Delta^2}$ But, loss is proportional to $(\frac{g^2}{\kappa\Delta})^2$.Considering $m\frac{g^2}{\kappa\Delta} \sim \frac{1}{\sqrt{\eta_r N}}$ we eventually end up with loss decreased by a factor of \textit{m}, i.e. $\frac{6}{m \pi \eta_r}\frac{(\kappa/2)^2}{(\kappa/2)^2+\Delta^2}$. Now even with $\frac{(\kappa/2)^2}{(\kappa/2)^2+\Delta^2} \sim 1$ we can achieve low loss simply by having a moderate value of \textit{m}. For $\eta_r \sim 1$ and \textit{m} = 10 we will have 20$\%$ loss, while having last factor as close to 1 as possible. Hence, we can have low loss with as small a detuning as the AC stark shift approximation allows us to. Also the phase shift is multiplied by \textit{m}. So, now it will be much easier to achieve the new phase shift condition $m\frac{g^2}{\kappa\Delta} \sim \frac{1}{\sqrt{\eta_r N}}$ for much more moderate values of $g, \kappa$ and N, especially with a small value of $\Delta$ allowed. This will also mean that now both the signal and the atoms are in resonance with the cavity. So, there will also be no need for dynamically detuning the cavity using piezo or other techniques. However, there are also going to be extra losses from the optical switches used in the multipass and espcially due to the mode matching of the cavity with the optical fiber. The main hurdle for this scheme is that cavity mode-matching is quite bad for most optical systems right now. In the nano-cavities of \citep{zhong2017nanophotonic}, the optimal coupling transmission achieved right now is 27$\%$. The multipass-cavity combination will only be useful if this value can be improved significantly (to 95$\%$ or higher).

\section{Conclusion}

In summary, we performed a detailed theoretical analysis for a cavity-enhanced non-destructive photonic qubit detector using an atomic ensemble and determined the necessary parameters for implementation of the scheme in a rare-earth ion doped crystal. A single-pass configuration such as in the proof-of-principle experiment of Ref. \citep{sinclair2016proposal} is unable to reach single-photon level sensitivity due to a tradeoff between phase shift and loss. This can be overcome by using a cavity. However, we showed that the presence of the cavity also introduces a significant complication because the phase shift acquires a dependence on the probe photon number, in addition to the desired dependence on the signal photon number. We analyzed this effect in detail to determine the final probe state, using the Husimi Q-representation in phase space and calculating the quadrature distributions, which allowed us to determine the success probability and error rate of the scheme as a function of various parameter values. We modeled the cavity-enhanced loss and estimated system parameters for Nd:YVO nanocavities as an example system toward implementation. For a successful implementation of the scheme a small transverse area, high finesse and relatively long cavity are needed. Although these values are not achievable in current systems, we think that they are within reach, given the recent rapid progress in coupling rare-earth ions to optical cavities. We thus hope that the present work will prepare the ground for the experimental realization of non-destructive photonic qubit detection in the not too distant future.

\section{Acknowledgments}

The authors acknowledge useful discussions with W.Tittel, C.Thiel, P.Barclay, T.Zhong, A.Faraon, N.Sinclair, N.Lauk, R.Ahlefeldt and S.Wein. SG acknowledges support through a Dean’s International Doctoral Recruitment Scholarship of the University of Calgary, an Alberta Innovates Technology Futures (AITF) Graduate Student Scholarship (GSS) and an Izaak Walton Killam Doctoral Scholarship. CS acknowledges support from the Natural Sciences and Engineering Research Council of Canada (NSERC).


%


\begin{thebibliography}{46}%
\makeatletter
\providecommand \@ifxundefined [1]{%
 \@ifx{#1\undefined}
}%
\providecommand \@ifnum [1]{%
 \ifnum #1\expandafter \@firstoftwo
 \else \expandafter \@secondoftwo
 \fi
}%
\providecommand \@ifx [1]{%
 \ifx #1\expandafter \@firstoftwo
 \else \expandafter \@secondoftwo
 \fi
}%
\providecommand \natexlab [1]{#1}%
\providecommand \enquote  [1]{``#1''}%
\providecommand \bibnamefont  [1]{#1}%
\providecommand \bibfnamefont [1]{#1}%
\providecommand \citenamefont [1]{#1}%
\providecommand \href@noop [0]{\@secondoftwo}%
\providecommand \href [0]{\begingroup \@sanitize@url \@href}%
\providecommand \@href[1]{\@@startlink{#1}\@@href}%
\providecommand \@@href[1]{\endgroup#1\@@endlink}%
\providecommand \@sanitize@url [0]{\catcode `\\12\catcode `\$12\catcode
  `\&12\catcode `\#12\catcode `\^12\catcode `\_12\catcode `\%12\relax}%
\providecommand \@@startlink[1]{}%
\providecommand \@@endlink[0]{}%
\providecommand \url  [0]{\begingroup\@sanitize@url \@url }%
\providecommand \@url [1]{\endgroup\@href {#1}{\urlprefix }}%
\providecommand \urlprefix  [0]{URL }%
\providecommand \Eprint [0]{\href }%
\providecommand \doibase [0]{http://dx.doi.org/}%
\providecommand \selectlanguage [0]{\@gobble}%
\providecommand \bibinfo  [0]{\@secondoftwo}%
\providecommand \bibfield  [0]{\@secondoftwo}%
\providecommand \translation [1]{[#1]}%
\providecommand \BibitemOpen [0]{}%
\providecommand \bibitemStop [0]{}%
\providecommand \bibitemNoStop [0]{.\EOS\space}%
\providecommand \EOS [0]{\spacefactor3000\relax}%
\providecommand \BibitemShut  [1]{\csname bibitem#1\endcsname}%
\let\auto@bib@innerbib\@empty
\bibitem [{\citenamefont {Nemoto}\ and\ \citenamefont
  {Munro}(2004)}]{nemoto2004nearly}%
  \BibitemOpen
  \bibfield  {author} {\bibinfo {author} {\bibfnamefont {K.}~\bibnamefont
  {Nemoto}}\ and\ \bibinfo {author} {\bibfnamefont {W.~J.}\ \bibnamefont
  {Munro}},\ }\href@noop {} {\bibfield  {journal} {\bibinfo  {journal}
  {Physical Review Letters}\ }\textbf {\bibinfo {volume} {93}},\ \bibinfo
  {pages} {250502} (\bibinfo {year} {2004})}\BibitemShut {NoStop}%
\bibitem [{\citenamefont {Boone}\ \emph {et~al.}(2015)\citenamefont {Boone},
  \citenamefont {Bourgoin}, \citenamefont {Meyer-Scott}, \citenamefont
  {Heshami}, \citenamefont {Jennewein},\ and\ \citenamefont
  {Simon}}]{boone2015entanglement}%
  \BibitemOpen
  \bibfield  {author} {\bibinfo {author} {\bibfnamefont {K.}~\bibnamefont
  {Boone}}, \bibinfo {author} {\bibfnamefont {J.-P.}\ \bibnamefont {Bourgoin}},
  \bibinfo {author} {\bibfnamefont {E.}~\bibnamefont {Meyer-Scott}}, \bibinfo
  {author} {\bibfnamefont {K.}~\bibnamefont {Heshami}}, \bibinfo {author}
  {\bibfnamefont {T.}~\bibnamefont {Jennewein}}, \ and\ \bibinfo {author}
  {\bibfnamefont {C.}~\bibnamefont {Simon}},\ }\href@noop {} {\bibfield
  {journal} {\bibinfo  {journal} {Physical Review A}\ }\textbf {\bibinfo
  {volume} {91}},\ \bibinfo {pages} {052325} (\bibinfo {year}
  {2015})}\BibitemShut {NoStop}%
\bibitem [{\citenamefont {Simon}(2017)}]{simon2017towards}%
  \BibitemOpen
  \bibfield  {author} {\bibinfo {author} {\bibfnamefont {C.}~\bibnamefont
  {Simon}},\ }\href@noop {} {\bibfield  {journal} {\bibinfo  {journal} {Nature
  Photonics}\ }\textbf {\bibinfo {volume} {11}},\ \bibinfo {pages} {678}
  (\bibinfo {year} {2017})}\BibitemShut {NoStop}%
\bibitem [{\citenamefont {Chang}\ \emph {et~al.}(2014)\citenamefont {Chang},
  \citenamefont {Vuleti{\'c}},\ and\ \citenamefont {Lukin}}]{chang2014quantum}%
  \BibitemOpen
  \bibfield  {author} {\bibinfo {author} {\bibfnamefont {D.~E.}\ \bibnamefont
  {Chang}}, \bibinfo {author} {\bibfnamefont {V.}~\bibnamefont {Vuleti{\'c}}},
  \ and\ \bibinfo {author} {\bibfnamefont {M.~D.}\ \bibnamefont {Lukin}},\
  }\href@noop {} {\bibfield  {journal} {\bibinfo  {journal} {Nature Photonics}\
  }\textbf {\bibinfo {volume} {8}},\ \bibinfo {pages} {685} (\bibinfo {year}
  {2014})}\BibitemShut {NoStop}%
\bibitem [{\citenamefont {Reiserer}\ \emph {et~al.}(2013)\citenamefont
  {Reiserer}, \citenamefont {Ritter},\ and\ \citenamefont
  {Rempe}}]{reiserer2013nondestructive}%
  \BibitemOpen
  \bibfield  {author} {\bibinfo {author} {\bibfnamefont {A.}~\bibnamefont
  {Reiserer}}, \bibinfo {author} {\bibfnamefont {S.}~\bibnamefont {Ritter}}, \
  and\ \bibinfo {author} {\bibfnamefont {G.}~\bibnamefont {Rempe}},\
  }\href@noop {} {\bibfield  {journal} {\bibinfo  {journal} {Science}\ }\textbf
  {\bibinfo {volume} {342}},\ \bibinfo {pages} {1349} (\bibinfo {year}
  {2013})}\BibitemShut {NoStop}%
\bibitem [{\citenamefont {Saffman}\ \emph {et~al.}(2010)\citenamefont
  {Saffman}, \citenamefont {Walker},\ and\ \citenamefont
  {M{\o}lmer}}]{saffman2010quantum}%
  \BibitemOpen
  \bibfield  {author} {\bibinfo {author} {\bibfnamefont {M.}~\bibnamefont
  {Saffman}}, \bibinfo {author} {\bibfnamefont {T.~G.}\ \bibnamefont {Walker}},
  \ and\ \bibinfo {author} {\bibfnamefont {K.}~\bibnamefont {M{\o}lmer}},\
  }\href@noop {} {\bibfield  {journal} {\bibinfo  {journal} {Reviews of Modern
  Physics}\ }\textbf {\bibinfo {volume} {82}},\ \bibinfo {pages} {2313}
  (\bibinfo {year} {2010})}\BibitemShut {NoStop}%
\bibitem [{\citenamefont {Schmidt}\ and\ \citenamefont
  {Imamoglu}(1996)}]{schmidt1996giant}%
  \BibitemOpen
  \bibfield  {author} {\bibinfo {author} {\bibfnamefont {H.}~\bibnamefont
  {Schmidt}}\ and\ \bibinfo {author} {\bibfnamefont {A.}~\bibnamefont
  {Imamoglu}},\ }\href@noop {} {\bibfield  {journal} {\bibinfo  {journal}
  {Optics Letters}\ }\textbf {\bibinfo {volume} {21}},\ \bibinfo {pages} {1936}
  (\bibinfo {year} {1996})}\BibitemShut {NoStop}%
\bibitem [{\citenamefont {Hosseini}\ \emph {et~al.}(2016)\citenamefont
  {Hosseini}, \citenamefont {Beck}, \citenamefont {Duan}, \citenamefont
  {Chen},\ and\ \citenamefont {Vuleti{\'c}}}]{hosseini2016partially}%
  \BibitemOpen
  \bibfield  {author} {\bibinfo {author} {\bibfnamefont {M.}~\bibnamefont
  {Hosseini}}, \bibinfo {author} {\bibfnamefont {K.~M.}\ \bibnamefont {Beck}},
  \bibinfo {author} {\bibfnamefont {Y.}~\bibnamefont {Duan}}, \bibinfo {author}
  {\bibfnamefont {W.}~\bibnamefont {Chen}}, \ and\ \bibinfo {author}
  {\bibfnamefont {V.}~\bibnamefont {Vuleti{\'c}}},\ }\href@noop {} {\bibfield
  {journal} {\bibinfo  {journal} {Physical Review Letters}\ }\textbf {\bibinfo
  {volume} {116}},\ \bibinfo {pages} {033602} (\bibinfo {year}
  {2016})}\BibitemShut {NoStop}%
\bibitem [{\citenamefont {Feizpour}\ \emph {et~al.}(2015)\citenamefont
  {Feizpour}, \citenamefont {Hallaji}, \citenamefont {Dmochowski},\ and\
  \citenamefont {Steinberg}}]{feizpour2015observation}%
  \BibitemOpen
  \bibfield  {author} {\bibinfo {author} {\bibfnamefont {A.}~\bibnamefont
  {Feizpour}}, \bibinfo {author} {\bibfnamefont {M.}~\bibnamefont {Hallaji}},
  \bibinfo {author} {\bibfnamefont {G.}~\bibnamefont {Dmochowski}}, \ and\
  \bibinfo {author} {\bibfnamefont {A.~M.}\ \bibnamefont {Steinberg}},\
  }\href@noop {} {\bibfield  {journal} {\bibinfo  {journal} {Nature Physics}\
  }\textbf {\bibinfo {volume} {11}},\ \bibinfo {pages} {905} (\bibinfo {year}
  {2015})}\BibitemShut {NoStop}%
\bibitem [{\citenamefont {Zhong}\ \emph
  {et~al.}(2017{\natexlab{a}})\citenamefont {Zhong}, \citenamefont {Kindem},
  \citenamefont {Bartholomew}, \citenamefont {Rochman}, \citenamefont
  {Craiciu}, \citenamefont {Miyazono}, \citenamefont {Bettinelli},
  \citenamefont {Cavalli}, \citenamefont {Verma}, \citenamefont {Nam} \emph
  {et~al.}}]{zhong2017nanophotonic}%
  \BibitemOpen
  \bibfield  {author} {\bibinfo {author} {\bibfnamefont {T.}~\bibnamefont
  {Zhong}}, \bibinfo {author} {\bibfnamefont {J.~M.}\ \bibnamefont {Kindem}},
  \bibinfo {author} {\bibfnamefont {J.~G.}\ \bibnamefont {Bartholomew}},
  \bibinfo {author} {\bibfnamefont {J.}~\bibnamefont {Rochman}}, \bibinfo
  {author} {\bibfnamefont {I.}~\bibnamefont {Craiciu}}, \bibinfo {author}
  {\bibfnamefont {E.}~\bibnamefont {Miyazono}}, \bibinfo {author}
  {\bibfnamefont {M.}~\bibnamefont {Bettinelli}}, \bibinfo {author}
  {\bibfnamefont {E.}~\bibnamefont {Cavalli}}, \bibinfo {author} {\bibfnamefont
  {V.}~\bibnamefont {Verma}}, \bibinfo {author} {\bibfnamefont {S.~W.}\
  \bibnamefont {Nam}},  \emph {et~al.},\ }\href@noop {} {\bibfield  {journal}
  {\bibinfo  {journal} {Science}\ }\textbf {\bibinfo {volume} {357}},\ \bibinfo
  {pages} {1392} (\bibinfo {year} {2017}{\natexlab{a}})}\BibitemShut {NoStop}%
\bibitem [{\citenamefont {Zhong}\ \emph
  {et~al.}(2017{\natexlab{b}})\citenamefont {Zhong}, \citenamefont {Kindem},
  \citenamefont {Rochman},\ and\ \citenamefont
  {Faraon}}]{zhong2017interfacing}%
  \BibitemOpen
  \bibfield  {author} {\bibinfo {author} {\bibfnamefont {T.}~\bibnamefont
  {Zhong}}, \bibinfo {author} {\bibfnamefont {J.~M.}\ \bibnamefont {Kindem}},
  \bibinfo {author} {\bibfnamefont {J.}~\bibnamefont {Rochman}}, \ and\
  \bibinfo {author} {\bibfnamefont {A.}~\bibnamefont {Faraon}},\ }\href@noop {}
  {\bibfield  {journal} {\bibinfo  {journal} {Nature communications}\ }\textbf
  {\bibinfo {volume} {8}} (\bibinfo {year} {2017}{\natexlab{b}})}\BibitemShut
  {NoStop}%
\bibitem [{\citenamefont {Bartholomew}\ \emph {et~al.}(2018)\citenamefont
  {Bartholomew}, \citenamefont {Zhong}, \citenamefont {Kindem}, \citenamefont
  {Lopez-Rios}, \citenamefont {Rochman}, \citenamefont {Craiciu}, \citenamefont
  {Miyazono},\ and\ \citenamefont {Faraon}}]{faraon2018controlling}%
  \BibitemOpen
  \bibfield  {author} {\bibinfo {author} {\bibfnamefont {J.~G.}\ \bibnamefont
  {Bartholomew}}, \bibinfo {author} {\bibfnamefont {T.}~\bibnamefont {Zhong}},
  \bibinfo {author} {\bibfnamefont {J.~M.}\ \bibnamefont {Kindem}}, \bibinfo
  {author} {\bibfnamefont {R.}~\bibnamefont {Lopez-Rios}}, \bibinfo {author}
  {\bibfnamefont {J.}~\bibnamefont {Rochman}}, \bibinfo {author} {\bibfnamefont
  {I.}~\bibnamefont {Craiciu}}, \bibinfo {author} {\bibfnamefont
  {E.}~\bibnamefont {Miyazono}}, \ and\ \bibinfo {author} {\bibfnamefont
  {A.}~\bibnamefont {Faraon}},\ }\href@noop {} {\bibfield  {journal} {\bibinfo
  {journal} {arXiv preprint arXiv:1802.06172}\ } (\bibinfo {year}
  {2018})}\BibitemShut {NoStop}%
\bibitem [{\citenamefont {Dibos}\ \emph {et~al.}(2017)\citenamefont {Dibos},
  \citenamefont {Raha}, \citenamefont {Phenicie},\ and\ \citenamefont
  {Thompson}}]{jeff2017isolating}%
  \BibitemOpen
  \bibfield  {author} {\bibinfo {author} {\bibfnamefont {A.}~\bibnamefont
  {Dibos}}, \bibinfo {author} {\bibfnamefont {M.}~\bibnamefont {Raha}},
  \bibinfo {author} {\bibfnamefont {C.}~\bibnamefont {Phenicie}}, \ and\
  \bibinfo {author} {\bibfnamefont {J.}~\bibnamefont {Thompson}},\ }\href@noop
  {} {\bibfield  {journal} {\bibinfo  {journal} {arXiv preprint
  arXiv:1711.10368}\ } (\bibinfo {year} {2017})}\BibitemShut {NoStop}%
\bibitem [{\citenamefont {Casabone}\ \emph {et~al.}(2018)\citenamefont
  {Casabone}, \citenamefont {Benedikter}, \citenamefont {H{\"u}mmer},
  \citenamefont {Beck}, \citenamefont {Lima}, \citenamefont {H{\"a}nsch},
  \citenamefont {Ferrier}, \citenamefont {Goldner}, \citenamefont
  {de~Riedmatten},\ and\ \citenamefont {Hunger}}]{hunger2018cavity}%
  \BibitemOpen
  \bibfield  {author} {\bibinfo {author} {\bibfnamefont {B.}~\bibnamefont
  {Casabone}}, \bibinfo {author} {\bibfnamefont {J.}~\bibnamefont
  {Benedikter}}, \bibinfo {author} {\bibfnamefont {T.}~\bibnamefont
  {H{\"u}mmer}}, \bibinfo {author} {\bibfnamefont {F.}~\bibnamefont {Beck}},
  \bibinfo {author} {\bibfnamefont {K.~d.~O.}\ \bibnamefont {Lima}}, \bibinfo
  {author} {\bibfnamefont {T.~W.}\ \bibnamefont {H{\"a}nsch}}, \bibinfo
  {author} {\bibfnamefont {A.}~\bibnamefont {Ferrier}}, \bibinfo {author}
  {\bibfnamefont {P.}~\bibnamefont {Goldner}}, \bibinfo {author} {\bibfnamefont
  {H.}~\bibnamefont {de~Riedmatten}}, \ and\ \bibinfo {author} {\bibfnamefont
  {D.}~\bibnamefont {Hunger}},\ }\href@noop {} {\bibfield  {journal} {\bibinfo
  {journal} {arXiv preprint arXiv:1802.06709}\ } (\bibinfo {year}
  {2018})}\BibitemShut {NoStop}%
\bibitem [{\citenamefont {O'Brien}\ \emph {et~al.}(2016)\citenamefont
  {O'Brien}, \citenamefont {Zhong}, \citenamefont {Faraon},\ and\ \citenamefont
  {Simon}}]{o2016nondestructive}%
  \BibitemOpen
  \bibfield  {author} {\bibinfo {author} {\bibfnamefont {C.}~\bibnamefont
  {O'Brien}}, \bibinfo {author} {\bibfnamefont {T.}~\bibnamefont {Zhong}},
  \bibinfo {author} {\bibfnamefont {A.}~\bibnamefont {Faraon}}, \ and\ \bibinfo
  {author} {\bibfnamefont {C.}~\bibnamefont {Simon}},\ }\href@noop {}
  {\bibfield  {journal} {\bibinfo  {journal} {Physical Review A}\ }\textbf
  {\bibinfo {volume} {94}},\ \bibinfo {pages} {043807} (\bibinfo {year}
  {2016})}\BibitemShut {NoStop}%
\bibitem [{\citenamefont {Chen}\ \emph {et~al.}(2013)\citenamefont {Chen},
  \citenamefont {Beck}, \citenamefont {B{\"u}cker}, \citenamefont {Gullans},
  \citenamefont {Lukin}, \citenamefont {Tanji-Suzuki},\ and\ \citenamefont
  {Vuleti{\'c}}}]{chen2013all}%
  \BibitemOpen
  \bibfield  {author} {\bibinfo {author} {\bibfnamefont {W.}~\bibnamefont
  {Chen}}, \bibinfo {author} {\bibfnamefont {K.~M.}\ \bibnamefont {Beck}},
  \bibinfo {author} {\bibfnamefont {R.}~\bibnamefont {B{\"u}cker}}, \bibinfo
  {author} {\bibfnamefont {M.}~\bibnamefont {Gullans}}, \bibinfo {author}
  {\bibfnamefont {M.~D.}\ \bibnamefont {Lukin}}, \bibinfo {author}
  {\bibfnamefont {H.}~\bibnamefont {Tanji-Suzuki}}, \ and\ \bibinfo {author}
  {\bibfnamefont {V.}~\bibnamefont {Vuleti{\'c}}},\ }\href@noop {} {\bibfield
  {journal} {\bibinfo  {journal} {Science}\ }\textbf {\bibinfo {volume}
  {341}},\ \bibinfo {pages} {768} (\bibinfo {year} {2013})}\BibitemShut
  {NoStop}%
\bibitem [{\citenamefont {Baur}\ \emph {et~al.}(2014)\citenamefont {Baur},
  \citenamefont {Tiarks}, \citenamefont {Rempe},\ and\ \citenamefont
  {D\"urr}}]{PhysRevLett.112.073901}%
  \BibitemOpen
  \bibfield  {author} {\bibinfo {author} {\bibfnamefont {S.}~\bibnamefont
  {Baur}}, \bibinfo {author} {\bibfnamefont {D.}~\bibnamefont {Tiarks}},
  \bibinfo {author} {\bibfnamefont {G.}~\bibnamefont {Rempe}}, \ and\ \bibinfo
  {author} {\bibfnamefont {S.}~\bibnamefont {D\"urr}},\ }\href {\doibase
  10.1103/PhysRevLett.112.073901} {\bibfield  {journal} {\bibinfo  {journal}
  {Physical Review Letters}\ }\textbf {\bibinfo {volume} {112}},\ \bibinfo
  {pages} {073901} (\bibinfo {year} {2014})}\BibitemShut {NoStop}%
\bibitem [{\citenamefont {Venkataraman}\ \emph {et~al.}(2013)\citenamefont
  {Venkataraman}, \citenamefont {Saha},\ and\ \citenamefont
  {Gaeta}}]{venkataraman2013phase}%
  \BibitemOpen
  \bibfield  {author} {\bibinfo {author} {\bibfnamefont {V.}~\bibnamefont
  {Venkataraman}}, \bibinfo {author} {\bibfnamefont {K.}~\bibnamefont {Saha}},
  \ and\ \bibinfo {author} {\bibfnamefont {A.~L.}\ \bibnamefont {Gaeta}},\
  }\href@noop {} {\bibfield  {journal} {\bibinfo  {journal} {Nature Photonics}\
  }\textbf {\bibinfo {volume} {7}},\ \bibinfo {pages} {138} (\bibinfo {year}
  {2013})}\BibitemShut {NoStop}%
\bibitem [{\citenamefont {Heshami}\ \emph {et~al.}(2016)\citenamefont
  {Heshami}, \citenamefont {England}, \citenamefont {Humphreys}, \citenamefont
  {Bustard}, \citenamefont {Acosta}, \citenamefont {Nunn},\ and\ \citenamefont
  {Sussman}}]{heshami2016quantum}%
  \BibitemOpen
  \bibfield  {author} {\bibinfo {author} {\bibfnamefont {K.}~\bibnamefont
  {Heshami}}, \bibinfo {author} {\bibfnamefont {D.~G.}\ \bibnamefont
  {England}}, \bibinfo {author} {\bibfnamefont {P.~C.}\ \bibnamefont
  {Humphreys}}, \bibinfo {author} {\bibfnamefont {P.~J.}\ \bibnamefont
  {Bustard}}, \bibinfo {author} {\bibfnamefont {V.~M.}\ \bibnamefont {Acosta}},
  \bibinfo {author} {\bibfnamefont {J.}~\bibnamefont {Nunn}}, \ and\ \bibinfo
  {author} {\bibfnamefont {B.~J.}\ \bibnamefont {Sussman}},\ }\href@noop {}
  {\bibfield  {journal} {\bibinfo  {journal} {Journal of Modern Optics}\
  }\textbf {\bibinfo {volume} {63}},\ \bibinfo {pages} {2005} (\bibinfo {year}
  {2016})}\BibitemShut {NoStop}%
\bibitem [{\citenamefont {Sinclair}\ \emph {et~al.}(2016)\citenamefont
  {Sinclair}, \citenamefont {Heshami}, \citenamefont {Deshmukh}, \citenamefont
  {Oblak}, \citenamefont {Simon},\ and\ \citenamefont
  {Tittel}}]{sinclair2016proposal}%
  \BibitemOpen
  \bibfield  {author} {\bibinfo {author} {\bibfnamefont {N.}~\bibnamefont
  {Sinclair}}, \bibinfo {author} {\bibfnamefont {K.}~\bibnamefont {Heshami}},
  \bibinfo {author} {\bibfnamefont {C.}~\bibnamefont {Deshmukh}}, \bibinfo
  {author} {\bibfnamefont {D.}~\bibnamefont {Oblak}}, \bibinfo {author}
  {\bibfnamefont {C.}~\bibnamefont {Simon}}, \ and\ \bibinfo {author}
  {\bibfnamefont {W.}~\bibnamefont {Tittel}},\ }\href@noop {} {\bibfield
  {journal} {\bibinfo  {journal} {Nature communications}\ }\textbf {\bibinfo
  {volume} {7}},\ \bibinfo {pages} {13454} (\bibinfo {year}
  {2016})}\BibitemShut {NoStop}%
\bibitem [{\citenamefont {Autler}\ and\ \citenamefont
  {Townes}(1955)}]{autler1955stark}%
  \BibitemOpen
  \bibfield  {author} {\bibinfo {author} {\bibfnamefont {S.~H.}\ \bibnamefont
  {Autler}}\ and\ \bibinfo {author} {\bibfnamefont {C.~H.}\ \bibnamefont
  {Townes}},\ }\href@noop {} {\bibfield  {journal} {\bibinfo  {journal}
  {Physical Review}\ }\textbf {\bibinfo {volume} {100}},\ \bibinfo {pages}
  {703} (\bibinfo {year} {1955})}\BibitemShut {NoStop}%
\bibitem [{\citenamefont {Yang}\ \emph {et~al.}(2016)\citenamefont {Yang},
  \citenamefont {Wang}, \citenamefont {Bao},\ and\ \citenamefont
  {Pan}}]{yang2016efficient}%
  \BibitemOpen
  \bibfield  {author} {\bibinfo {author} {\bibfnamefont {S.-J.}\ \bibnamefont
  {Yang}}, \bibinfo {author} {\bibfnamefont {X.-J.}\ \bibnamefont {Wang}},
  \bibinfo {author} {\bibfnamefont {X.-H.}\ \bibnamefont {Bao}}, \ and\
  \bibinfo {author} {\bibfnamefont {J.-W.}\ \bibnamefont {Pan}},\ }\href@noop
  {} {\bibfield  {journal} {\bibinfo  {journal} {Nature Photonics}\ }\textbf
  {\bibinfo {volume} {10}},\ \bibinfo {pages} {381} (\bibinfo {year}
  {2016})}\BibitemShut {NoStop}%
\bibitem [{\citenamefont {Afzelius}\ \emph {et~al.}(2009)\citenamefont
  {Afzelius}, \citenamefont {Simon}, \citenamefont {De~Riedmatten},\ and\
  \citenamefont {Gisin}}]{afzelius2009multimode}%
  \BibitemOpen
  \bibfield  {author} {\bibinfo {author} {\bibfnamefont {M.}~\bibnamefont
  {Afzelius}}, \bibinfo {author} {\bibfnamefont {C.}~\bibnamefont {Simon}},
  \bibinfo {author} {\bibfnamefont {H.}~\bibnamefont {De~Riedmatten}}, \ and\
  \bibinfo {author} {\bibfnamefont {N.}~\bibnamefont {Gisin}},\ }\href@noop {}
  {\bibfield  {journal} {\bibinfo  {journal} {Physical Review A}\ }\textbf
  {\bibinfo {volume} {79}},\ \bibinfo {pages} {052329} (\bibinfo {year}
  {2009})}\BibitemShut {NoStop}%
\bibitem [{\citenamefont {De~Riedmatten}\ \emph {et~al.}(2008)\citenamefont
  {De~Riedmatten}, \citenamefont {Afzelius}, \citenamefont {Staudt},
  \citenamefont {Simon},\ and\ \citenamefont {Gisin}}]{de2008solid}%
  \BibitemOpen
  \bibfield  {author} {\bibinfo {author} {\bibfnamefont {H.}~\bibnamefont
  {De~Riedmatten}}, \bibinfo {author} {\bibfnamefont {M.}~\bibnamefont
  {Afzelius}}, \bibinfo {author} {\bibfnamefont {M.~U.}\ \bibnamefont
  {Staudt}}, \bibinfo {author} {\bibfnamefont {C.}~\bibnamefont {Simon}}, \
  and\ \bibinfo {author} {\bibfnamefont {N.}~\bibnamefont {Gisin}},\
  }\href@noop {} {\bibfield  {journal} {\bibinfo  {journal} {Nature}\ }\textbf
  {\bibinfo {volume} {456}},\ \bibinfo {pages} {773} (\bibinfo {year}
  {2008})}\BibitemShut {NoStop}%
\bibitem [{\citenamefont {Hedges}\ \emph {et~al.}(2010)\citenamefont {Hedges},
  \citenamefont {Longdell}, \citenamefont {Li},\ and\ \citenamefont
  {Sellars}}]{hedges2010efficient}%
  \BibitemOpen
  \bibfield  {author} {\bibinfo {author} {\bibfnamefont {M.~P.}\ \bibnamefont
  {Hedges}}, \bibinfo {author} {\bibfnamefont {J.~J.}\ \bibnamefont
  {Longdell}}, \bibinfo {author} {\bibfnamefont {Y.}~\bibnamefont {Li}}, \ and\
  \bibinfo {author} {\bibfnamefont {M.~J.}\ \bibnamefont {Sellars}},\
  }\href@noop {} {\bibfield  {journal} {\bibinfo  {journal} {Nature}\ }\textbf
  {\bibinfo {volume} {465}},\ \bibinfo {pages} {1052} (\bibinfo {year}
  {2010})}\BibitemShut {NoStop}%
\bibitem [{\citenamefont {Zhong}\ \emph
  {et~al.}(2015{\natexlab{a}})\citenamefont {Zhong}, \citenamefont {Hedges},
  \citenamefont {Ahlefeldt}, \citenamefont {Bartholomew}, \citenamefont
  {Beavan}, \citenamefont {Wittig}, \citenamefont {Longdell},\ and\
  \citenamefont {Sellars}}]{zhong2015optically}%
  \BibitemOpen
  \bibfield  {author} {\bibinfo {author} {\bibfnamefont {M.}~\bibnamefont
  {Zhong}}, \bibinfo {author} {\bibfnamefont {M.~P.}\ \bibnamefont {Hedges}},
  \bibinfo {author} {\bibfnamefont {R.~L.}\ \bibnamefont {Ahlefeldt}}, \bibinfo
  {author} {\bibfnamefont {J.~G.}\ \bibnamefont {Bartholomew}}, \bibinfo
  {author} {\bibfnamefont {S.~E.}\ \bibnamefont {Beavan}}, \bibinfo {author}
  {\bibfnamefont {S.~M.}\ \bibnamefont {Wittig}}, \bibinfo {author}
  {\bibfnamefont {J.~J.}\ \bibnamefont {Longdell}}, \ and\ \bibinfo {author}
  {\bibfnamefont {M.~J.}\ \bibnamefont {Sellars}},\ }\href@noop {} {\bibfield
  {journal} {\bibinfo  {journal} {Nature}\ }\textbf {\bibinfo {volume} {517}},\
  \bibinfo {pages} {177} (\bibinfo {year} {2015}{\natexlab{a}})}\BibitemShut
  {NoStop}%
\bibitem [{\citenamefont {Gorshkov}\ \emph {et~al.}(2007)\citenamefont
  {Gorshkov}, \citenamefont {Andr{\'e}}, \citenamefont {Fleischhauer},
  \citenamefont {S{\o}rensen},\ and\ \citenamefont
  {Lukin}}]{gorshkov2007universal}%
  \BibitemOpen
  \bibfield  {author} {\bibinfo {author} {\bibfnamefont {A.~V.}\ \bibnamefont
  {Gorshkov}}, \bibinfo {author} {\bibfnamefont {A.}~\bibnamefont {Andr{\'e}}},
  \bibinfo {author} {\bibfnamefont {M.}~\bibnamefont {Fleischhauer}}, \bibinfo
  {author} {\bibfnamefont {A.~S.}\ \bibnamefont {S{\o}rensen}}, \ and\ \bibinfo
  {author} {\bibfnamefont {M.~D.}\ \bibnamefont {Lukin}},\ }\href@noop {}
  {\bibfield  {journal} {\bibinfo  {journal} {Physical Review Letters}\
  }\textbf {\bibinfo {volume} {98}},\ \bibinfo {pages} {123601} (\bibinfo
  {year} {2007})}\BibitemShut {NoStop}%
\bibitem [{\citenamefont {Sabooni}\ \emph {et~al.}(2013)\citenamefont
  {Sabooni}, \citenamefont {Li}, \citenamefont {Kr{\"o}ll},\ and\ \citenamefont
  {Rippe}}]{sabooni2013efficient}%
  \BibitemOpen
  \bibfield  {author} {\bibinfo {author} {\bibfnamefont {M.}~\bibnamefont
  {Sabooni}}, \bibinfo {author} {\bibfnamefont {Q.}~\bibnamefont {Li}},
  \bibinfo {author} {\bibfnamefont {S.}~\bibnamefont {Kr{\"o}ll}}, \ and\
  \bibinfo {author} {\bibfnamefont {L.}~\bibnamefont {Rippe}},\ }\href@noop {}
  {\bibfield  {journal} {\bibinfo  {journal} {Physical Review Letters}\
  }\textbf {\bibinfo {volume} {110}},\ \bibinfo {pages} {133604} (\bibinfo
  {year} {2013})}\BibitemShut {NoStop}%
\bibitem [{\citenamefont {Saunders}\ \emph {et~al.}(2016)\citenamefont
  {Saunders}, \citenamefont {Munns}, \citenamefont {Champion}, \citenamefont
  {Qiu}, \citenamefont {Kaczmarek}, \citenamefont {Poem}, \citenamefont
  {Ledingham}, \citenamefont {Walmsley},\ and\ \citenamefont
  {Nunn}}]{saunders2016cavity}%
  \BibitemOpen
  \bibfield  {author} {\bibinfo {author} {\bibfnamefont {D.}~\bibnamefont
  {Saunders}}, \bibinfo {author} {\bibfnamefont {J.}~\bibnamefont {Munns}},
  \bibinfo {author} {\bibfnamefont {T.}~\bibnamefont {Champion}}, \bibinfo
  {author} {\bibfnamefont {C.}~\bibnamefont {Qiu}}, \bibinfo {author}
  {\bibfnamefont {K.}~\bibnamefont {Kaczmarek}}, \bibinfo {author}
  {\bibfnamefont {E.}~\bibnamefont {Poem}}, \bibinfo {author} {\bibfnamefont
  {P.}~\bibnamefont {Ledingham}}, \bibinfo {author} {\bibfnamefont
  {I.}~\bibnamefont {Walmsley}}, \ and\ \bibinfo {author} {\bibfnamefont
  {J.}~\bibnamefont {Nunn}},\ }\href@noop {} {\bibfield  {journal} {\bibinfo
  {journal} {Physical Review Letters}\ }\textbf {\bibinfo {volume} {116}},\
  \bibinfo {pages} {090501} (\bibinfo {year} {2016})}\BibitemShut {NoStop}%
\bibitem [{\citenamefont {Reim}\ \emph {et~al.}(2010)\citenamefont {Reim},
  \citenamefont {Nunn}, \citenamefont {Lorenz}, \citenamefont {Sussman},
  \citenamefont {Lee}, \citenamefont {Langford}, \citenamefont {Jaksch},\ and\
  \citenamefont {Walmsley}}]{reim2010towards}%
  \BibitemOpen
  \bibfield  {author} {\bibinfo {author} {\bibfnamefont {K.}~\bibnamefont
  {Reim}}, \bibinfo {author} {\bibfnamefont {J.}~\bibnamefont {Nunn}}, \bibinfo
  {author} {\bibfnamefont {V.}~\bibnamefont {Lorenz}}, \bibinfo {author}
  {\bibfnamefont {B.}~\bibnamefont {Sussman}}, \bibinfo {author} {\bibfnamefont
  {K.}~\bibnamefont {Lee}}, \bibinfo {author} {\bibfnamefont {N.}~\bibnamefont
  {Langford}}, \bibinfo {author} {\bibfnamefont {D.}~\bibnamefont {Jaksch}}, \
  and\ \bibinfo {author} {\bibfnamefont {I.}~\bibnamefont {Walmsley}},\
  }\href@noop {} {\bibfield  {journal} {\bibinfo  {journal} {Nature Photonics}\
  }\textbf {\bibinfo {volume} {4}},\ \bibinfo {pages} {218} (\bibinfo {year}
  {2010})}\BibitemShut {NoStop}%
\bibitem [{\citenamefont {Zhong}\ \emph
  {et~al.}(2015{\natexlab{b}})\citenamefont {Zhong}, \citenamefont {Kindem},
  \citenamefont {Miyazono},\ and\ \citenamefont
  {Faraon}}]{zhong2015nanophotonic}%
  \BibitemOpen
  \bibfield  {author} {\bibinfo {author} {\bibfnamefont {T.}~\bibnamefont
  {Zhong}}, \bibinfo {author} {\bibfnamefont {J.~M.}\ \bibnamefont {Kindem}},
  \bibinfo {author} {\bibfnamefont {E.}~\bibnamefont {Miyazono}}, \ and\
  \bibinfo {author} {\bibfnamefont {A.}~\bibnamefont {Faraon}},\ }\href@noop {}
  {\bibfield  {journal} {\bibinfo  {journal} {Nature communications}\ }\textbf
  {\bibinfo {volume} {6}} (\bibinfo {year} {2015}{\natexlab{b}})}\BibitemShut
  {NoStop}%
\bibitem [{\citenamefont {Zhong}\ \emph {et~al.}(2016)\citenamefont {Zhong},
  \citenamefont {Rochman}, \citenamefont {Kindem}, \citenamefont {Miyazono},\
  and\ \citenamefont {Faraon}}]{zhong2016high}%
  \BibitemOpen
  \bibfield  {author} {\bibinfo {author} {\bibfnamefont {T.}~\bibnamefont
  {Zhong}}, \bibinfo {author} {\bibfnamefont {J.}~\bibnamefont {Rochman}},
  \bibinfo {author} {\bibfnamefont {J.~M.}\ \bibnamefont {Kindem}}, \bibinfo
  {author} {\bibfnamefont {E.}~\bibnamefont {Miyazono}}, \ and\ \bibinfo
  {author} {\bibfnamefont {A.}~\bibnamefont {Faraon}},\ }\href@noop {}
  {\bibfield  {journal} {\bibinfo  {journal} {Optics express}\ }\textbf
  {\bibinfo {volume} {24}},\ \bibinfo {pages} {536} (\bibinfo {year}
  {2016})}\BibitemShut {NoStop}%
\bibitem [{\citenamefont {Ahlefeldt}\ \emph {et~al.}(2013)\citenamefont
  {Ahlefeldt}, \citenamefont {McAuslan}, \citenamefont {Longdell},
  \citenamefont {Manson},\ and\ \citenamefont
  {Sellars}}]{ahlefeldt2013precision}%
  \BibitemOpen
  \bibfield  {author} {\bibinfo {author} {\bibfnamefont {R.}~\bibnamefont
  {Ahlefeldt}}, \bibinfo {author} {\bibfnamefont {D.}~\bibnamefont {McAuslan}},
  \bibinfo {author} {\bibfnamefont {J.}~\bibnamefont {Longdell}}, \bibinfo
  {author} {\bibfnamefont {N.}~\bibnamefont {Manson}}, \ and\ \bibinfo {author}
  {\bibfnamefont {M.}~\bibnamefont {Sellars}},\ }\href@noop {} {\bibfield
  {journal} {\bibinfo  {journal} {Physical Review Letters}\ }\textbf {\bibinfo
  {volume} {111}},\ \bibinfo {pages} {240501} (\bibinfo {year}
  {2013})}\BibitemShut {NoStop}%
\bibitem [{\citenamefont {Ahlefeldt}\ \emph {et~al.}(2016)\citenamefont
  {Ahlefeldt}, \citenamefont {Hush},\ and\ \citenamefont
  {Sellars}}]{ahlefeldt2016ultranarrow}%
  \BibitemOpen
  \bibfield  {author} {\bibinfo {author} {\bibfnamefont {R.}~\bibnamefont
  {Ahlefeldt}}, \bibinfo {author} {\bibfnamefont {M.}~\bibnamefont {Hush}}, \
  and\ \bibinfo {author} {\bibfnamefont {M.}~\bibnamefont {Sellars}},\
  }\href@noop {} {\bibfield  {journal} {\bibinfo  {journal} {Physical review
  letters}\ }\textbf {\bibinfo {volume} {117}},\ \bibinfo {pages} {250504}
  (\bibinfo {year} {2016})}\BibitemShut {NoStop}%
\bibitem [{\citenamefont {Scully}\ and\ \citenamefont
  {Zubairy}(1997)}]{scully1999quantum}%
  \BibitemOpen
  \bibfield  {author} {\bibinfo {author} {\bibfnamefont {M.~O.}\ \bibnamefont
  {Scully}}\ and\ \bibinfo {author} {\bibfnamefont {M.~S.}\ \bibnamefont
  {Zubairy}},\ }\href@noop {} {\emph {\bibinfo {title} {Quantum Optics}}}\
  (\bibinfo  {publisher} {Cambridge University Press},\ \bibinfo {year}
  {1997})\BibitemShut {NoStop}%
\bibitem [{\citenamefont {Duan}\ and\ \citenamefont
  {Kimble}(2004)}]{duan2004scalable}%
  \BibitemOpen
  \bibfield  {author} {\bibinfo {author} {\bibfnamefont {L.-M.}\ \bibnamefont
  {Duan}}\ and\ \bibinfo {author} {\bibfnamefont {H.}~\bibnamefont {Kimble}},\
  }\href@noop {} {\bibfield  {journal} {\bibinfo  {journal} {Physical Review
  Letters}\ }\textbf {\bibinfo {volume} {92}},\ \bibinfo {pages} {127902}
  (\bibinfo {year} {2004})}\BibitemShut {NoStop}%
\bibitem [{\citenamefont {Husimi}(1940)}]{husimi1940some}%
  \BibitemOpen
  \bibfield  {author} {\bibinfo {author} {\bibfnamefont {K.}~\bibnamefont
  {Husimi}},\ }\href@noop {} {\bibfield  {journal} {\bibinfo  {journal}
  {Proceedings of the Physico-Mathematical Society of Japan. 3rd Series}\
  }\textbf {\bibinfo {volume} {22}},\ \bibinfo {pages} {264} (\bibinfo {year}
  {1940})}\BibitemShut {NoStop}%
\bibitem [{\citenamefont {Wolfowicz}\ \emph {et~al.}(2015)\citenamefont
  {Wolfowicz}, \citenamefont {Maier-Flaig}, \citenamefont {Marino},
  \citenamefont {Ferrier}, \citenamefont {Vezin}, \citenamefont {Morton},\ and\
  \citenamefont {Goldner}}]{wolfowicz2015coherent}%
  \BibitemOpen
  \bibfield  {author} {\bibinfo {author} {\bibfnamefont {G.}~\bibnamefont
  {Wolfowicz}}, \bibinfo {author} {\bibfnamefont {H.}~\bibnamefont
  {Maier-Flaig}}, \bibinfo {author} {\bibfnamefont {R.}~\bibnamefont {Marino}},
  \bibinfo {author} {\bibfnamefont {A.}~\bibnamefont {Ferrier}}, \bibinfo
  {author} {\bibfnamefont {H.}~\bibnamefont {Vezin}}, \bibinfo {author}
  {\bibfnamefont {J.~J.}\ \bibnamefont {Morton}}, \ and\ \bibinfo {author}
  {\bibfnamefont {P.}~\bibnamefont {Goldner}},\ }\href@noop {} {\bibfield
  {journal} {\bibinfo  {journal} {Physical Review Letters}\ }\textbf {\bibinfo
  {volume} {114}},\ \bibinfo {pages} {170503} (\bibinfo {year}
  {2015})}\BibitemShut {NoStop}%
\bibitem [{\citenamefont {Afzelius}\ \emph {et~al.}(2010)\citenamefont
  {Afzelius}, \citenamefont {Staudt}, \citenamefont {De~Riedmatten},
  \citenamefont {Gisin}, \citenamefont {Guillot-No{\"e}l}, \citenamefont
  {Goldner}, \citenamefont {Marino}, \citenamefont {Porcher}, \citenamefont
  {Cavalli},\ and\ \citenamefont {Bettinelli}}]{afzelius2010efficient}%
  \BibitemOpen
  \bibfield  {author} {\bibinfo {author} {\bibfnamefont {M.}~\bibnamefont
  {Afzelius}}, \bibinfo {author} {\bibfnamefont {M.~U.}\ \bibnamefont
  {Staudt}}, \bibinfo {author} {\bibfnamefont {H.}~\bibnamefont
  {De~Riedmatten}}, \bibinfo {author} {\bibfnamefont {N.}~\bibnamefont
  {Gisin}}, \bibinfo {author} {\bibfnamefont {O.}~\bibnamefont
  {Guillot-No{\"e}l}}, \bibinfo {author} {\bibfnamefont {P.}~\bibnamefont
  {Goldner}}, \bibinfo {author} {\bibfnamefont {R.}~\bibnamefont {Marino}},
  \bibinfo {author} {\bibfnamefont {P.}~\bibnamefont {Porcher}}, \bibinfo
  {author} {\bibfnamefont {E.}~\bibnamefont {Cavalli}}, \ and\ \bibinfo
  {author} {\bibfnamefont {M.}~\bibnamefont {Bettinelli}},\ }\href@noop {}
  {\bibfield  {journal} {\bibinfo  {journal} {Journal of Luminescence}\
  }\textbf {\bibinfo {volume} {130}},\ \bibinfo {pages} {1566} (\bibinfo {year}
  {2010})}\BibitemShut {NoStop}%
\bibitem [{\citenamefont {Hastings-Simon}\ \emph {et~al.}(2008)\citenamefont
  {Hastings-Simon}, \citenamefont {Afzelius}, \citenamefont {Min{\'a}{\v{r}}},
  \citenamefont {Staudt}, \citenamefont {Lauritzen}, \citenamefont
  {de~Riedmatten}, \citenamefont {Gisin}, \citenamefont {Amari}, \citenamefont
  {Walther}, \citenamefont {Kr{\"o}ll} \emph {et~al.}}]{hastings2008spectral}%
  \BibitemOpen
  \bibfield  {author} {\bibinfo {author} {\bibfnamefont {S.}~\bibnamefont
  {Hastings-Simon}}, \bibinfo {author} {\bibfnamefont {M.}~\bibnamefont
  {Afzelius}}, \bibinfo {author} {\bibfnamefont {J.}~\bibnamefont
  {Min{\'a}{\v{r}}}}, \bibinfo {author} {\bibfnamefont {M.}~\bibnamefont
  {Staudt}}, \bibinfo {author} {\bibfnamefont {B.}~\bibnamefont {Lauritzen}},
  \bibinfo {author} {\bibfnamefont {H.}~\bibnamefont {de~Riedmatten}}, \bibinfo
  {author} {\bibfnamefont {N.}~\bibnamefont {Gisin}}, \bibinfo {author}
  {\bibfnamefont {A.}~\bibnamefont {Amari}}, \bibinfo {author} {\bibfnamefont
  {A.}~\bibnamefont {Walther}}, \bibinfo {author} {\bibfnamefont
  {S.}~\bibnamefont {Kr{\"o}ll}},  \emph {et~al.},\ }\href@noop {} {\bibfield
  {journal} {\bibinfo  {journal} {Physical Review B}\ }\textbf {\bibinfo
  {volume} {77}},\ \bibinfo {pages} {125111} (\bibinfo {year}
  {2008})}\BibitemShut {NoStop}%
\bibitem [{\citenamefont {Choi}\ \emph {et~al.}(2017)\citenamefont {Choi},
  \citenamefont {Heuck},\ and\ \citenamefont {Englund}}]{englund2017self}%
  \BibitemOpen
  \bibfield  {author} {\bibinfo {author} {\bibfnamefont {H.}~\bibnamefont
  {Choi}}, \bibinfo {author} {\bibfnamefont {M.}~\bibnamefont {Heuck}}, \ and\
  \bibinfo {author} {\bibfnamefont {D.}~\bibnamefont {Englund}},\ }\href@noop
  {} {\bibfield  {journal} {\bibinfo  {journal} {Physical Review Letters}\
  }\textbf {\bibinfo {volume} {118}},\ \bibinfo {pages} {223605} (\bibinfo
  {year} {2017})}\BibitemShut {NoStop}%
\bibitem [{\citenamefont {Peng}\ \emph {et~al.}(2001)\citenamefont {Peng},
  \citenamefont {Asundi}, \citenamefont {Chen},\ and\ \citenamefont
  {Xiong}}]{peng2001study}%
  \BibitemOpen
  \bibfield  {author} {\bibinfo {author} {\bibfnamefont {X.}~\bibnamefont
  {Peng}}, \bibinfo {author} {\bibfnamefont {A.}~\bibnamefont {Asundi}},
  \bibinfo {author} {\bibfnamefont {Y.}~\bibnamefont {Chen}}, \ and\ \bibinfo
  {author} {\bibfnamefont {Z.}~\bibnamefont {Xiong}},\ }\href@noop {}
  {\bibfield  {journal} {\bibinfo  {journal} {Applied Optics}\ }\textbf
  {\bibinfo {volume} {40}},\ \bibinfo {pages} {1396} (\bibinfo {year}
  {2001})}\BibitemShut {NoStop}%
\bibitem [{Win()}]{WinNT}%
  \BibitemOpen
  \href@noop {} {\enquote {\bibinfo {title} {Pi cermaic actuators},}\ }\bibinfo
  {howpublished}
  {\url{https://www.piceramic.com/en/products/piezoceramic-actuators/linear-actuators/}}\BibitemShut
  {NoStop}%
\bibitem [{\citenamefont {Yang}\ \emph {et~al.}(2004)\citenamefont {Yang},
  \citenamefont {Lee}, \citenamefont {Mueller},\ and\ \citenamefont
  {George}}]{yang2004leak}%
  \BibitemOpen
  \bibfield  {author} {\bibinfo {author} {\bibfnamefont {E.-H.}\ \bibnamefont
  {Yang}}, \bibinfo {author} {\bibfnamefont {C.}~\bibnamefont {Lee}}, \bibinfo
  {author} {\bibfnamefont {J.}~\bibnamefont {Mueller}}, \ and\ \bibinfo
  {author} {\bibfnamefont {T.}~\bibnamefont {George}},\ }\href@noop {}
  {\bibfield  {journal} {\bibinfo  {journal} {Journal of Microelectromechanical
  Systems}\ }\textbf {\bibinfo {volume} {13}},\ \bibinfo {pages} {799}
  (\bibinfo {year} {2004})}\BibitemShut {NoStop}%
\bibitem [{\citenamefont {Judy}\ \emph {et~al.}(2009)\citenamefont {Judy},
  \citenamefont {Pulskamp}, \citenamefont {Polcawich},\ and\ \citenamefont
  {Currano}}]{judy2009piezoelectric}%
  \BibitemOpen
  \bibfield  {author} {\bibinfo {author} {\bibfnamefont {D.}~\bibnamefont
  {Judy}}, \bibinfo {author} {\bibfnamefont {J.}~\bibnamefont {Pulskamp}},
  \bibinfo {author} {\bibfnamefont {R.}~\bibnamefont {Polcawich}}, \ and\
  \bibinfo {author} {\bibfnamefont {L.}~\bibnamefont {Currano}},\ }in\
  \href@noop {} {\emph {\bibinfo {booktitle} {Micro Electro Mechanical Systems,
  2009. MEMS 2009. IEEE 22nd International Conference on}}}\ (\bibinfo
  {organization} {IEEE},\ \bibinfo {year} {2009})\ pp.\ \bibinfo {pages}
  {591--594}\BibitemShut {NoStop}%
\bibitem [{\citenamefont {Hua}\ and\ \citenamefont
  {Vohra}(1997)}]{hua1997pressure}%
  \BibitemOpen
  \bibfield  {author} {\bibinfo {author} {\bibfnamefont {H.}~\bibnamefont
  {Hua}}\ and\ \bibinfo {author} {\bibfnamefont {Y.~K.}\ \bibnamefont
  {Vohra}},\ }\href@noop {} {\bibfield  {journal} {\bibinfo  {journal} {Applied
  physics letters}\ }\textbf {\bibinfo {volume} {71}},\ \bibinfo {pages} {2602}
  (\bibinfo {year} {1997})}\BibitemShut {NoStop}%
\end{thebibliography}
\end{document}